\documentclass{aa}
\usepackage{graphics}

\newcommand{\comment}[1]{}
        \def\smallskip{\vskip 2pt}
\input{psfig.sty}

\begin{document}

\thesaurus{06(08.05.3; 08.16.4; 08.13.2; 09.16.1; 09.01.1; 11.01.1)}

\title{Chemical yields from low- and intermediate-mass stars: \\
       model predictions and basic observational constraints} 

\author{Paola Marigo}
\institute{Dipartimento di Astronomia, Universit\`a di Padova,
	Vicolo dell'Osservatorio 2, I-35122 Padova, Italia}

\offprints{Paola Marigo \\ e-mail: marigo@pd.astro.it} 

\date{Received 14 September 2000 / Accepted}

\maketitle
\markboth{P. Marigo }{Yields from low- and intermediate-mass stars}

\begin{abstract}
In this work we analyse the role of low- and intermediate-mass stars
in contributing to the chemical enrichment of the interstellar medium.
First we present new sets of stellar yields basing on 
the results of updated evolutionary calculations, which extend from
the ZAMS up to the end of the AGB phase 
(Girardi et al. 2000; Marigo et al. 1999a).
These new yields, that present a significant dependence on metallicity, 
are then compared to those of other available sets  
(Renzini \& Voli 1981; van de Hoek \& Groenewegen 1997).
The resulting differences are explained in terms of different model 
assumptions  
-- i.e. treatment of convective boundaries, 
mass loss, dredge-up, hot-bottom burning -- 
 and further discussed on the basis of important empirical
constraints which should be reproduced by theory
-- i.e. the initial-final mass relation, white dwarf mass distribution,
carbon star luminosity function, and chemical abundances of planetary nebulae.
We show that present models are able to reproduce such constraints in a 
satisfactory way. 

\end{abstract}
\section{Introduction}
\label{intro}
Low- and intermediate-mass stars 
(with masses 0.8 $M_{\odot} \la M \la 5-8 M_{\odot}$, depending 
on model details)
play an important role in galactic
chemical evolution, thanks to the ejection into the interstellar medium (ISM) 
of material containing newly synthesized nuclear products, mainly
$^{4}$He, $^{12}$C, and $^{14}$N, and possibly $^{16}$O.

The interest in the nucleosynthetic history of $^{12}$C and $^{14}$N, 
in particular, has  recently increased thanks to the observations of
high red-shift systems. For instance, 
it is possible to measure the nitrogen abundance
and the N/O ratio in the damped Ly-$\alpha$ systems 
(e.g. Pettini et al. 1995; Lu et al. 1998), 
whereas carbon is detected in the Ly-$\alpha$ forest clouds
(e.g. Lu 1991; Crotts et al. 1994, Tytler  \& Fan 1994).

The chemical history of both these elements in a galaxy 
is quite complex, as it depends, besides other factors, 
on the relative contributions of stars with different 
masses, hence releasing their yields at different timescales.
In the case of $^{12}$C the
relative role of low- and intermediate-mass  stars,
or rather of metal-rich, mass-losing massive stars, in 
the carbon enrichment at late galactic epochs is still a matter
of debate (Prantzos et al. 1994).
In the case of $^{14}$N, large uncertainties affect 
the theoretical predictions on both the secondary and primary nucleosynthesis
of nitrogen in low- and intermediate-mass stars, and 
the possible contribution of primary nitrogen from massive stars
(Woosley \& Weaver 1982, Maeder 1983; see also the review by 
Maeder \& Meynet 2000).

It follows that, in order to trace the history of chemical evolution
and of star formation in the Universe -- which the observations of high
red-shift systems are intended to infer -- it is quite important to analyse in
detail the nucleosynthesis of these elements taking place in  low- and
intermediate-mass stars.
The contribution of these stars to the chemical enrichment
of the ISM essentially occurs during the RGB and AGB phases, 
both characterized by the occurrence of mass loss and 
events of surface chemical pollution (dredge-up episodes).

Whereas modelling the evolution of a star on the RGB 
is rather easy by means of modern stellar evolutionary codes,
dealing with the most advanced stages -- characterised by the occurrence 
of thermal pulses (the TP-AGB phase) -- is quite a difficult task, 
due to both the high complexity of the physics involved, and
the remarkable requirement of computing time.
Then, an alternative theoretical approach is offered 
by {\sl synthetic} models, which summarise 
the results of {\sl complete} stellar calculations through simple 
and practical analytical relations. This allows quick computing of the models 
and ready analysis of the results.

Historically, the first most significant   
AGB synthetic models were those developed   
by Iben \& Truran (1978) and Renzini \& Voli (1981).
These pioneer works focused on the 
contribution of low- and intermediate-mass stars to the 
element enrichment of the ISM. In particular,  
the sets of stellar yields presented by Renzini \& Voli (1981)
have been extensively used in chemical evolution models 
of galaxies, providing  almost the only available data source up 
to recent years.

It is important to remark that in Renzini \& Voli (1981) 
synthetic AGB model 
the fundamental parameters (essentially mass-loss and dredge-up) were
specified according to the indications from the current knowledge 
of the involved physical processes and available complete stellar calculations.
In this sense the model was {\sl uncalibrated}.
That approach had the merit of supplying a testing tool for complete
stellar models, as it
pointed out at some fundamental inadequacies 
in the predictions (e.g. too low efficiency
of the third dredge-up), and assumed prescriptions (e.g. too low
mass-loss rates), which led to clear discrepancies between theory
and observations (see, for instance, Iben 1981; Iben \& Renzini 1983; 
Bragaglia et al. 1995). 

With awareness of that,
the later synthetic AGB models 
(e.g. Groenewegen \& de Jong 1993; Marigo et al. 1996, 1999a) 
have made the next step ahead, that is update the input prescriptions 
and {\sl calibrate} the model parameters in order
to reproduce fundamental observables 
(e.g. the carbon star luminosity functions, the initial-final mass
relation). 
Basing on the results of these model calibrations, 
various sets of stellar yields 
from low- and intermediate-mass stars have been presented in recent years
as an alternative to Renzini \& Voli (1981), 
namely: Marigo et al. (1996, 1998), van de Hoek \& Groenewegen (1997);
see also Forestini \& Charbonnel (1997), and  Boothroyd
\& Sackmann (1999). 

To the above reference list we will now add the  
new homogeneous sets of stellar yields presented in this work.
With respect to our previous calculations (Marigo et al. 1996, 1998),
the present yields are derived from stellar models with updated input 
prescriptions and improved treatment of the relevant processes involved
(i.e. the third dredge-up, see Marigo et al. 1999a for all details).

Specifically, we follow the evolution of low- and intermediate-mass stars,
coupling the results of complete stellar models (Girardi et al. 2000) -- 
that cover the evolution from the zero age main sequence (ZAMS) up 
to the onset of the thermally pulsing asymptotic giant branch (TP-AGB) --
with synthetic TP-AGB models (Marigo 1998ab; Marigo et al. 1999a) 
that extend the calculations up to the end of this phase.
Relevant model prescriptions are briefly recalled in Sect.~\ref{models}.
    
With the aid of these evolutionary calculations,
we then derive the stellar yields (H, He, and main CNO elements; refer
to Sects.~\ref{yi} and \ref{stelyie})  
for a dense grid of initial stellar masses 
(in the range $0.8 M_{\odot}$ -- $5 M_{\odot}$) and 
various metallicities ($Z=$0.004, 0.008, 0.019). 
For the most massive stars, 
experiencing hot-bottom burning (hereinafter HBB, or envelope burning) 
during the TP-AGB phase, the 
corresponding yields are given for  three values of the 
mixing-length parameter, i.e. $\alpha=1.68$, 
$\alpha=2.00$, and $\alpha=2.50$.
This parameter mainly affects the predicted production 
of $^{14}$N and $^{4}$He due to HBB (See Sect.~\ref{primy}).

The final part of this work (Sect.~\ref{compare_yields}) 
is dedicated to compare 
our results with the yields calculated by other
authors, i.e. Renzini \& Voli (1981), and van de Hoek \& Groenewegen (1997).
In the attempt to single out the causes of the main differences, we 
analyse  
the effect of different model prescriptions (e.g. mass-loss, dredge-up, 
HBB) on the predicted yields, testing at the same time 
the capability of a model to satisfy basic observational constraints 
(e.g. initial-final mass relations, white dwarf mass
distribution, carbon star luminosity functions, chemical abundances of 
planetary nebulae).

\section{Evolutionary models}
\label{models}
\subsection{Some definitions}
\label{somedef}
Let us first define some quantities which will be often used throughout
this paper to indicate critical masses 
for the occurrence of particular physical processes.
These quantities are: $M_{\rm HeF}$, 
$M_{\rm up}$, $M_{\rm HBB}^{\rm min}$, $M_{\rm dred}^{\rm min}$, and 
$M_{\rm c}^{\rm min}$.

The first one, $M_{\rm HeF}$, 
denotes the maximum initial mass for a star to develop a degenerate He-core, 
hence experience the He-flash at the tip of the Red Gian Branch (RGB), and
is comprised between 1.7 -- 2.5 $M_{\odot}$ 
depending on metallicity and model details.

The second  one, $M_{\rm up}$, 
is defined as the critical stellar mass over which 
carbon ignition occurs in non-degenerate conditions, marking the 
boundary between intermediate-mass and massive stars.
It is worth recalling that this mass limit 
is usually comprised within 5 -- 8 $M_{\odot}$, being significantly affected by
the adopted treatment of convective boundaries, as discussed 
in Sect.~\ref{compare_yields}.

The third one, $M_{\rm HBB}^{\rm min}$, corresponds to
the minimum initial mass for a star to undergo HBB
during the TP-AGB phase. Stellar evolution calculations indicate that
$M_{\rm i} \ga M_{\rm HBB}^{\rm min} \sim 3.5 - 4.5 M_{\odot}$,
depending on metallicity (see, for instance,  Marigo 1998b).

The fourth one, $M_{\rm dred}^{\rm min}$,  denotes the minimum initial
mass for a star to experience the third dredge-up during the
TP-AGB phase. Observations of carbon stars suggest that
$M_{\rm dred}^{\rm min} \sim 1.1 - 1.5 M_{\odot}$, 
decreasing with the metallicity (see, for instance, Marigo et al. 1999a).

Finally, we recall the basic parameters of the third dredge-up:
$\lambda$ and $M_{\rm c}^{\rm min}$. 
The parameter $\lambda$ is intended to measure   
the efficiency the third dredge-up, being defined as the 
fraction of the increment of core mass over an inter-pulse period
which is dredged-up to the surface at the subsequent thermal pulse.
The actual values for $\lambda$ are still a matter of debate among
theoreticians, and can range from $\lambda \sim 0$ to  $\lambda \ga 1$,
depending both on stellar properties (e.g. mass and metallicity) and 
model details (e.g. treatment of convective boundaries).

The parameter $M_{\rm c}^{\rm min}$ refers to the minimum core mass
for the occurrence of the third dredge-up, and is affected by
several factors as well.
In general, theoretical models would predict that $M_{\rm c}^{\rm min}$
decreases -- so that the third dredge-up is favoured -- at increasing
mass and mixing-length parameter, 
and decreasing metallicity (see, for instance, Wood 1981; Marigo et al. 1999a).

\subsection{Input physics}
In this work we consider 
low- and intermediate-mass stars, i.e. those  
with initial masses in the range from about 
$0.8 M_{\odot}$ to $M_{\rm up}$
$\sim 5.0 M_{\odot}$ and 
and three choices of the original compositions (i.e. $[Y=0.273, Z=0.019]$, 
$[Y=0.250, Z=0.008]$, $[Y=0.240, Z=0.004]$).

Their evolution from the ZAMS up to the beginning of the TP-AGB phase 
is taken from
the Padua stellar models (Girardi et al. 2000), 
that include moderate overshoot 
from core and external convection (Chiosi et al. 1992; Alongi et al. 1993).
The reader should refer to Girardi et al. (2000) for more
details of the adopted input physics.

The models by Girardi et al. (2000) provide the expected changes 
in the surface abundance of several chemical elements 
(H, $^3$He, $^4$He, $^{12}$C,
$^{13}$C, $^{14}$N, $^{15}$N, $^{16}$O, $^{17}$O,$^{18}$O, $^{20}$Ne,
$^{22}$Ne, $^{25}$Mg) caused by the first and second dredge-up episodes,
i.e. prior to the onset of the TP-AGB phase.
It should be also remarked that these models do not assume any ad-hoc 
``extra-mixing mechanism'',
e.g. the so-called cool-bottom process (Wasserburg et al. 1995; 
see also Charbonnel 1995; Boothroyd \& Sackmann 1999; Weiss et al. 2000),
which is invoked to reconcile discrepant predictions of surface 
abundances with  those measured in field Population II stars,  
galactic globular clusters, and Magellanic Clouds clusters 
(see, for instance, the results by Gratton et al. 2000 in their 
chemical analysis of field metal-poor stars). 

Mass loss by stellar winds suffered by low-mass stars 
(with $M_{\rm i} \le M_{\rm HeF} \sim 1.7-2.2 M_{\odot}$) on the ascent 
of RGB,  
is analytically included applying the
classical Reimers (1975) formula to the evolutionary tracks calculated 
at constant mass by Girardi et al. (2000).
An efficiency parameter $\eta = 0.45$ is adopted to fulfil the classical
observational constraint provided by the morphology of horizontal branches
in Galactic Globular Clusters.

Finally, once the
the first significant thermal pulse is singled out in each evolutionary
sequence calculated by Girardi et al. (2000), 
that point is assumed to define the starting 
conditions for synthetic calculations of the TP-AGB phase
(following the model prescriptions described in  Marigo 1998ab, and Marigo et
al. 1999a), which are carried on up to the complete ejection
of the envelope by stellar winds.
Mass loss is included according to the semi-empirical formalism
developed by Vassiliadis \& Wood (1993).
Nucleosynthesis occurring in the innermost envelope layers of TP-AGB stars
with HBB is followed adopting the Caughlan \& Fowler 
(1988) compilation of reaction rates for CNO and p-p reactions.
The electron-screening factors are those of Graboske et al. (1973).

\section{Derivation of the wind contributions}
\label{yi}
Following the classical definition by Tinsley (1980),
 the {\it stellar yield}, $p_{k}(M_{\rm i})$,
of a given chemical element $k$, is the mass fraction of a star with initial
mass $M_{\rm i}$ that is converted into the element $k$ and returned to the 
ISM during its entire lifetime, $\tau(M_{\rm i})$.
Tables A1 -- A12 give the 
quantities:
\begin{equation}
M_{\rm y}(k) = M_{\rm i}\, p_{k}(M_{\rm i})
\label{defy1}
\end{equation}
expressed in solar masses,
for all the chemical elements considered, as function of the initial stellar
mass $M_{\rm i}$,  metallicity $Z$, and mixing-length parameter $\alpha$.

According to the definition of stellar yield we can write:
\begin{equation}
M_{\rm y}(k) = \int_{0}^{\tau(M_{\rm i})} [X(k)-X^{0}(k)] 
\,\, \frac{{\rm d}M}{{\rm d} t} {\rm d}t
\label{defy}
\end{equation}
where ${\rm d}M/{\rm d t}$ is the current mass-loss rate;
$X(k)$ and $X^{0}(k)$ refer to the current
and initial surface abundance of the element $k$, respectively.

Thanks to the fact that the surface chemical composition of stars 
not suffering HBB during the TP-AGB phase (i.e. with initial
masses $M_{\rm i} \la M_{\rm HBB}^{\rm min}$),
is altered by the occurrence of discrete 
and quasi-instantaneous episodes of convective
dredge-up, that alternate with periods of continuous mass-loss, 
the evaluation of 
stellar yields can be simplified as follows.

Denoting by $X^1(k)$, $X^2(k)$, $X^j(k)$,
the abundance
of the species $k$ after the first
dredge-up, the second dredge-up, 
and the $j^{\rm th}$ dredge-up event during the TP-AGB phase, 
respectively, 
we get:
\begin{equation}
M_{\rm y}(k) = M_{\rm y}(k)_{\rm RGB} + M_{\rm y}(k)_{\rm E-AGB} +
               M_{\rm y}(k)_{\rm TP-AGB}
\label{totyield}
\end{equation}
where
\begin{equation}
M_{\rm y}(k)_{\rm RGB} = [X^1(k) - X^0(k)]\,\, \Delta M^{\rm ej}_{\rm RGB}
\label{RGBml}
\end{equation}
\begin{equation}
M_{\rm y}(k)_{\rm E-AGB} = [X^2(k) - X^0(k)]\,\, \Delta M^{\rm ej}_{\rm E-AGB}
\label{EAGBml}
\end{equation}
\begin{equation}
M_{\rm y}(k)_{\rm TP-AGB} = \sum_j [X^j(k) - X^0(k)]\,\,
\Delta M^{j, {\rm ej}}_{\rm TP-AGB}
\label{TPAGBml}
\end{equation}

In Eq.~(\ref{RGBml}) $\Delta M^{\rm ej}_{\rm RGB}$ is the mass of the envelope
ejected during the entire RGB phase (defined only for low-mass stars).
It is worth noticing that most of $\Delta M^{\rm ej}_{\rm RGB}$ is 
lost close to the tip of the RGB, that is 
after the occurrence of the first dredge-up, so that Eq.~(\ref{RGBml})
is a good approximation of Eq.~(\ref{defy}).

As far as mass loss on the AGB is concerned, 
we remark that, with the adopted prescription for $\dot M$
(Vassiliadis \& Wood 1993), 
the amount mass lost during the E-AGB phase is indeed negligible
so that we can assume $\Delta M^{\rm ej}_{\rm E-AGB} = 0$ 
in Eq.~(\ref{EAGBml}).

The contribution of the TP-AGB phase is evaluated with Eq.~(\ref{TPAGBml}),
that sums all the partial contributions of the pulse cycles, the generic 
$j^{\rm th}$ one  
consisting of a thermal pulse -- when the $j^{\rm th}$ dredge-up 
possibly  occurs --  followed by the inter-pulse period, during
which the mass $\Delta M^{j,{\rm ej}}_{\rm TP-AGB}$ is  ejected.

It should be noticed that this approximation holds for TP-AGB stars which 
experience only the third dredge-up. 
For more massive TP-AGB stars (with $M_{\rm i} > M_{\rm HBB}^{\rm min}$) 
also suffering HBB, the changes in the surface chemical composition  
and mass loss are concomitant processes, so that
the calculation of stellar yields requires 
the adoption of integration time steps shorter 
than the inter-pulse periods.

%
%

In general,  {\it negative} $M_{\rm y}(k)$
correspond to those elemental species which are prevalently
{\it destroyed} and diluted in the envelope,
 so that their abundances in the ejected
material are lower with respect to the main sequence values. 
On the contrary, {\it positive} $M_{\rm y}(k)$
correspond to those elements which are prevalently {\it produced}
so that a net enrichment of their abundances in the ejecta is predicted. 

\begin{figure*}
\resizebox{\hsize}{!}{\includegraphics{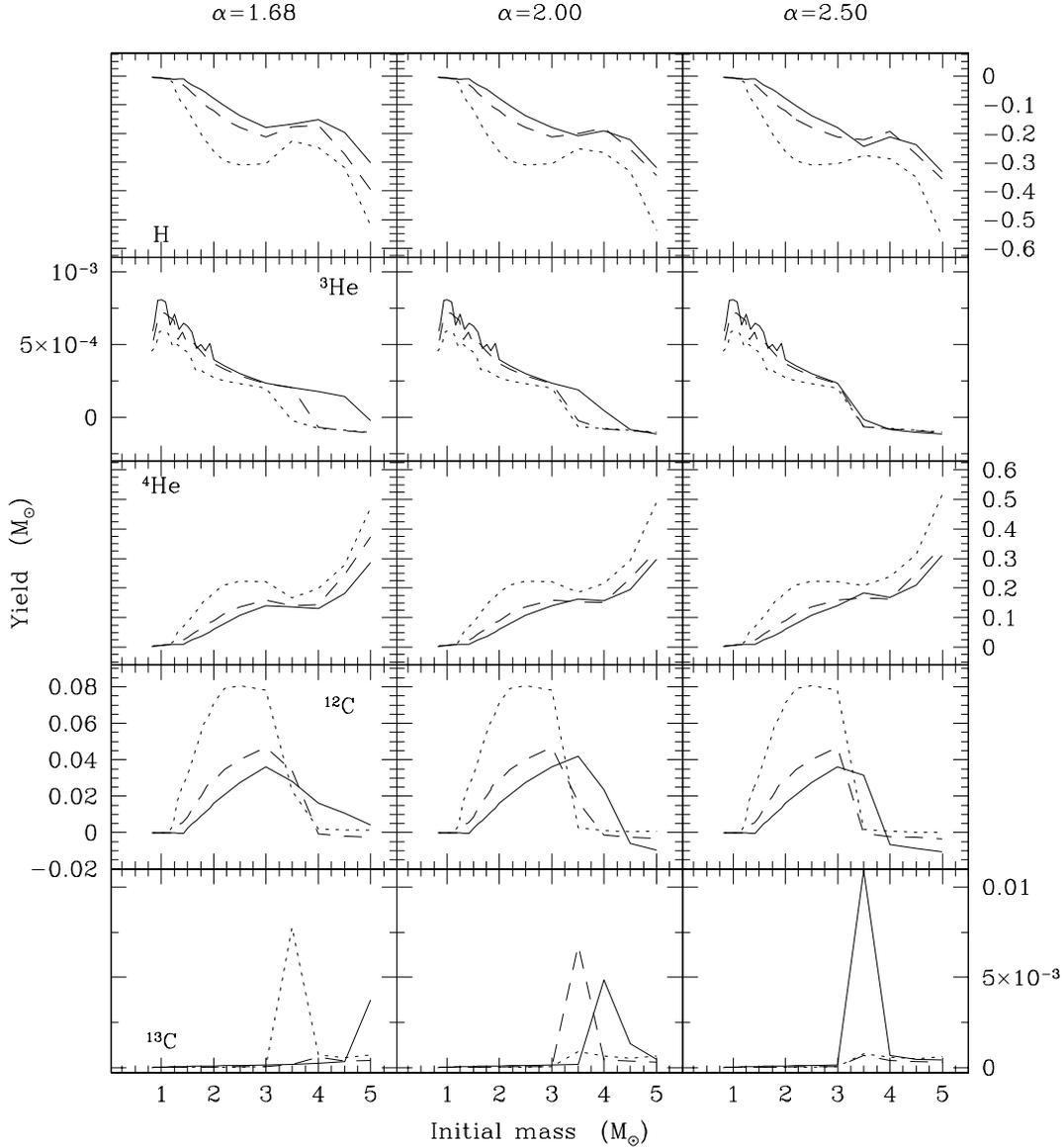}}
\caption{Net yields $M_{\rm y}(k)$ (in $M_{\odot}$) for each
indicated chemical element (H, $^{3}$He, $^{4}$He, $^{12}$C, and  $^{13}$C) 
as a function of the initial mass (in $M_{\odot}$) of the star.
The solid, dashed, and dotted lines correspond to the metallicity
sets $Z=0.019$, $Z=0.008$, and $Z=0.004$, respectively.
Panels along each column refer to the same value of the mixing-length
parameter.}
\label{stary1}
\end{figure*}

\begin{figure*}
\resizebox{\hsize}{!}{\includegraphics{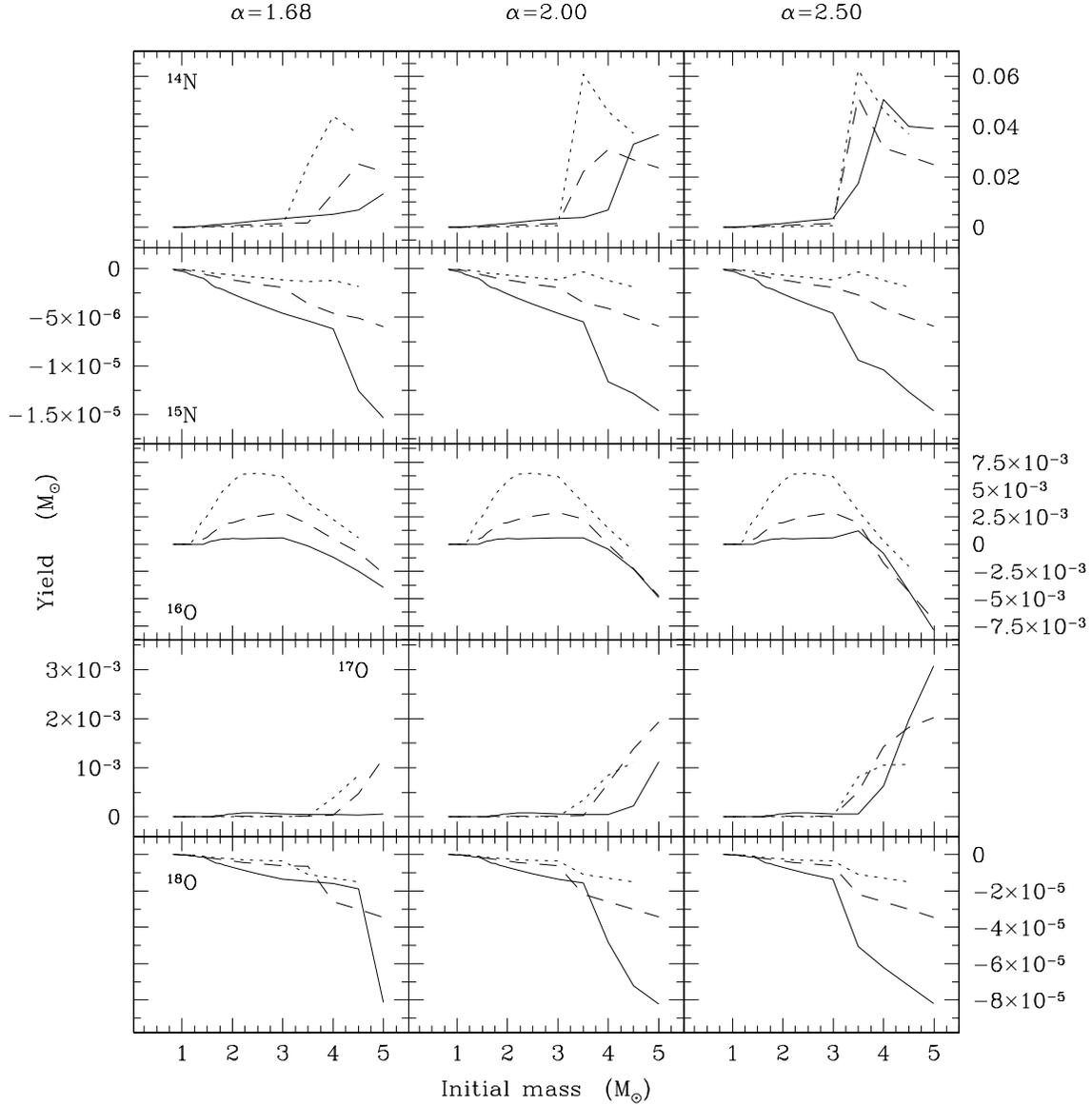}}
\caption{Net stellar yields $M_{\rm y}(k)$ (in $M_{\odot}$) for each
indicated chemical element ($^{14}$N, $^{15}$N, $^{16}$O, $^{17}$O, 
and $^{18}$O)
as a function of the initial stellar mass (in $M_{\odot}$).
The notation is the same as in Fig.~\protect\ref{stary1}.}
\label{stary2}
\end{figure*}

Figures~\ref{stary1} and \ref{stary2} show 
the quantities $M_{\rm y}(k)$ for all 
the chemical
elements considered, as a function of $M_{\rm i}$ and $Z$.
For stars experiencing HBB
results are given for three values of the mixing-length parameter.

Finally, for the sake of clarity,
we remind that the CNO cycle does not change 
the total number of CNO nuclei involved as 
catalysts in the conversion of H into $^4$He, i.e.
$Y_{\rm CNO} = \sum_{k} X_{k}^{\rm CNO}/A_{k} = constant$.
It follows that the first and second dredge-up, though affecting  the surface
abundances of the CNO isotopes, do not alter their {\sl total abundance by
number}. In fact, the material injected into the envelope has
experienced the CNO cycle involving only isotopes 
already present in the original composition.
On the contrary, the constancy of $Y_{\rm CNO}$ breaks down as soon as 
the dredge-up of primary carbon and oxygen, produced 
by $\alpha$-capture reactions at thermal pulses, occurs.

\begin{figure*}
\resizebox{\hsize}{!}{\includegraphics{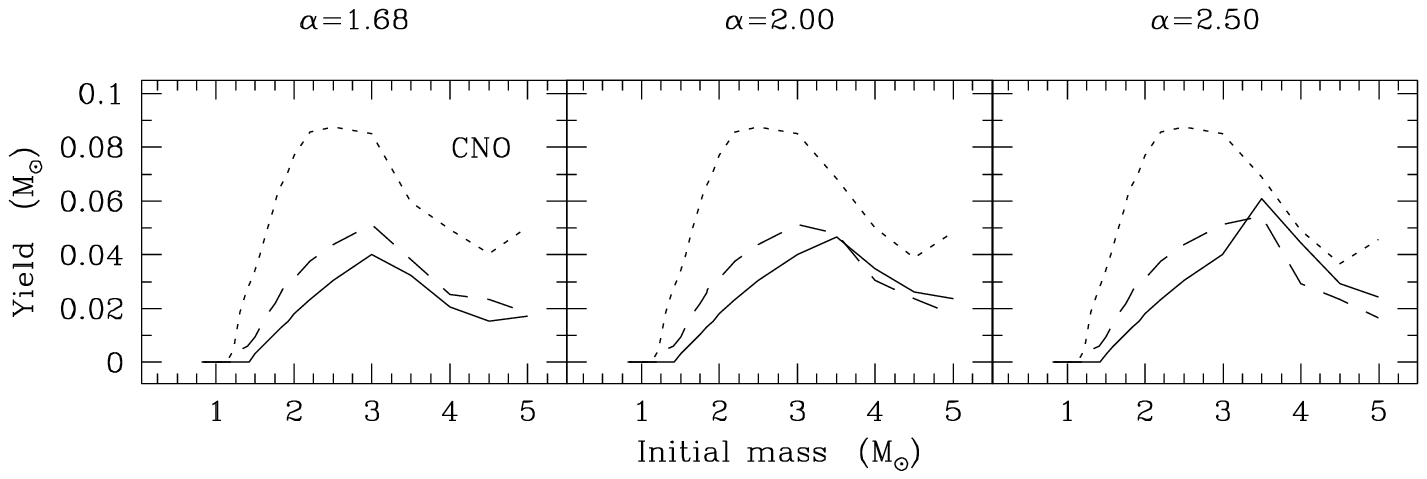}}
\caption{Net stellar yields of all CNO elements
as a function of the initial stellar mass.
The notation is the same as in Fig.~\protect\ref{stary1}.}
\label{stary3}
\end{figure*}

However, in all cases the {\sl total abundance by mass} of the CNO isotopes is 
somewhat changed because of the conversion of these elements mainly 
into $^{14}$N, so that
a small positive CNO yield (in mass fraction), is expected, for example,
from the RGB phase (see Tables A1 -- A3).
The quantity, 
$M_{\rm y}({\rm CNO})$, referring to
the total net yield of all CNO isotopes, 
is shown in Fig.~\ref{stary3}.

\section{Stellar yields as a function of $M_{\rm i}$, $Z$, and $\alpha$}
\label{stelyie}

Model predictions can be understood more easily considering that
the stellar yield of a given element is essentially determined by
the {\sl efficiency} and {\sl duration/frequency} of:
\begin{itemize}
\item the {\sl nucleosynthesis/mixing processes} 
(e.g. dredge-up events, HBB) which alter its abundance 
in the surface layers;
\item the {\sl mass-loss process} which ejects the surface layers
into the ISM
\end{itemize}

According to the physical prescriptions adopted in this work for the
TP-AGB phase (see also Sect.~\ref{modpre}) we can summarise 
the following points:
\begin{itemize}
\item The third dredge-up determines the surface enrichment mainly
of $^{4}$He, $^{12}$C, and $^{16}$O. 
The adopted intershell abundances are
[X($^{4}$He)$=0.76$;  X($^{12}$C)$=0.22$; X($^{4}$He)$=0.02$]
according to Boothroyd \& Sackmann (1988b).
The process is more efficient (i.e. higher $\lambda$) at lower $Z$;
\item HBB operates via the CNO-cycle, hence essentially increasing
the surface abundances of $^{4}$He, and $^{14}$N. The 
process is more efficient at higher $M$ (provided
that $M > M_{\rm HBB}^{\rm min}$), lower $Z$, and larger $\alpha$.
\item mass loss is, in general,
less efficient (i.e. lower $\dot M$) at decreasing $Z$ and 
increasing $\alpha$. In fact, both factors tend to produce hotter tracks,
and generally $\dot M$ anticorrelates with $T_{\rm eff}$.
As a consequence, lower mass-loss rates correspond to longer TP-AGB
lifetimes, hence greater number of 
third dredge-up events and a longer duration of HBB. 
\end{itemize}

The expected trend of $M_{\rm y}(k$) for the elements under consideration
as a function of the stellar initial mass, metallicity, and mixing-length
parameter is shown in Figs.~\ref{stary1}-\ref{stary3}. 
We can notice the following:

\begin{itemize}
\item {\bf H and $^{4}$He} \\
The net yields of these elements have mirror-like trends,
being negative for H, and positive for $^{4}$He.
The maximum of $^{4}$He  production at around $2-3 M_{\odot}$
(depending on metallicity) is 
explained considering the effect of 
the increase of the number of thermal pulses (hence dredge-up episodes)
with stellar mass for $ 0.8 M_{\odot} < M_{\rm i} \la 2-3 M_{\odot} $
(see Fig.~\ref{fig-npul}).
This peak is more pronounced at lower metallicities due to 
longer TP-AGB duration 
(Fig.~\ref{fig-npul}) and larger number of thermal pulses (Fig.~\ref{fig-time})
for given stellar mass.
The subsequent increase of $^{4}$He production towards higher masses,
$4 M_{\odot} \la M_{\rm i} \le 5 M_{\odot}$, is caused by the occurrence
of HBB in addition to the third dredge-up. 
The yield of  $^{4}$He
is larger for  lower metallicities
and higher values of the mixing-length parameter, reflecting the 
greater efficiency of both the third dredge-up and HBB.

\item {\bf $^{3}$He} \\
The net yield of this element presents a pronounced
peak at very low masses, say at $M_{\rm i} \sim1 M_{\odot}$,
and decreases at higher masses.
This trend is explained considering that 
the main contribution to the yield is due to the first 
dredge-up, and that both the related surface enrichment
of $^{3}$He and the 
amount of mass
lost during the RGB phase are inversely proportional to
the stellar mass in the low-mass domain.
At higher masses, ($M_{\rm i} \ga 3 M_{\odot}$), the net yield
 becomes even slightly negative because of HBB.
\item {\bf $^{12}$C} \\
The net positive yield increases with the initial mass,
up to a maximum located at  about $2-3 M_{\odot}$,
corresponding to largest number of dredge-up episodes
suffered during the TP-AGB phase, provided that HBB
has not operated.

The subsequent decline towards higher masses is
initially due to fewer dredge-up events, 
and then to the prevailing effect of HBB. 
It follows that no substantial enrichment of $^{12}$C is provided 
from the most massive AGB stars. 
 
This general trend is more marked at lower metallicities, because of the
longer TP-AGB phases, and the greater efficiency of the third dredge-up
and HBB.

\item {\bf $^{13}$C} \\
The first dredge-up causes an increase of the $^{13}$C surface  abundance
in stars of all masses, and the resulting yields are positive.
A further contribution is provided by mild HBB, as long as
creation of $^{13}$C via the reaction $^{12}$C$(p,\gamma)^{13}$C 
prevails over destruction via the reaction $^{13}$C$(p, \gamma)^{14}$N.
The results also depend on the interplay between
the strength  of HBB and mass loss.

The most favourable cases correspond to the 
models [e.g. ($4 M_{\odot}, Z=0.019, \alpha=2.00$), 
($3.5 M_{\odot}, Z=0.004, \alpha=1.68$)], in which the 
efficiency of reactions allows the synthesis of $^{13}$C 
for a long time before the drastic reduction
of the envelope causes the extinction of nuclear burning.
The spikes of $^{13}$C production 
for these models would suggest that 
a proper tuning of HBB is required: 
If nuclear reactions are  somewhat too weak $^{13}$C 
is not significantly created, 
else if somewhat too strong 
$^{13}$C is quickly destroyed in favour of $^{14}$N.

\item {\bf $^{14}$N} \\
It turns out that HBB plays the dominant role for the synthesis
of $^{14}$N. The positive yield as a function of the stellar mass depends on
both the efficiency and duration of nuclear burning. It follows that lower
metallicities and higher values of the mixing-length parameter concur to
favour nitrogen production. The contribution from low- and intermediate-mass
stars to the galactic enrichment of nitrogen may result important, as
suggested by chemical evolutionary models 
of galaxies (e.g. Portinari et al. 1998).

\item {\bf $^{15}$N} \\
The net yield of this element is mostly negative.
For stars with initial masses 
in the range, $0.8 M_{\odot} \la M_{\rm i} \la 3.5 M_{\odot}$,
the depletion of $^{15}$N is due to the effect of the first and second dredge-up.
For higher mass stars, the results are affected by HBB,
depending on the degree of CNO cycling attained in the burning regions.
This element has the shortest nuclear lifetime, after that of $^{18}$O,
so that it quickly attains nuclear equilibrium with $^{14}$N.
We note that the depletion of $^{15}$N is much more efficient at
higher  metallicities due to the increase of the CNO-cycling,
which implies a more efficient destruction of this element.
The predicted trend of the $^{15}$N yield mirrors, to some extent,
that of the secondary component of $^{14}$N (cf. Sect.~\ref{primy}).

\item {\bf $^{16}$O} \\
The net yield of this element is positive for stars with 
$0.8 M_{\odot} \la M_{\rm i} \la 3.5 M_{\odot}$ thanks to the dredge-up
events during the TP-AGB phase, whereas it becomes negative at higher
masses because of HBB.

The synthesis of fresh $^{16}$O occurs via the reaction 
$^{12}$C($\alpha,\gamma$)$^{16}$O during thermal pulses,
so that the yield of this element depends, among other factors, 
on its abundance in the dredged-up material.
According to recent TP-AGB calculations -- which include 
deep overshooting from all convective boundaries --
Herwig et al. (1997) find that the $^{16}$O  abundance in the 
convective intershell is roughly ten times higher, X($^{16}$O)$ \sim 0.25$,
than previously predicted, i.e. X($^{12}$C)$ \sim 0.02$ 
by Boothroyd \& Sackmann 1988b (standard case).
Then, adopting Herwig et al. indications, oxygen production
by low- and intermediate-mass stars may be favoured with respect
to the standard case, 
but it is worth recalling that, in general,
the final yields are crucially affected by various other factors
(e.g. number and efficiency of dredge-up episodes; 
see also Sect.~\ref{conclu}). 
 
Anyhow, regardless of its abundance in the dredged-up material, 
the trend of $^{16}$O, as a function of $M$ and $Z$, 
is expected to be qualitatively similar to that of $^{12}$C.
Moreover, according to the present calculations, 
a notable dependence on metallicity comes out.
The increasing positive trend with decreasing metallicities 
essentially reflects
the longer duration of the TP-AGB phase,  
and hence the greater number of dredge-up
episodes, in combination with their larger efficiency.
\item {\bf $^{17}$O} \\
A certain production of this element is provided by TP-AGB stars with
HBB, as long as the chain of reactions
$^{16}$O$(p,\gamma)^{17}$F$(\beta^{+} \nu)^{17}$O prevails over nuclear
destruction via the  
reactions $^{17}$O$(p,\gamma)^{18}$F and 
$^{17}$O$(p,\alpha)^{14}$N.
The behaviour resembles that of $^{13}$C, in the sense that under
some fine-tuned conditions -- for particular combinations 
of the stellar mass and metallicity -- a production spike of $^{17}$O
can result.

\item{\bf $^{18}$O} \\
This element has the shortest nuclear lifetime against proton captures,
so that it easily burns even at mild temperatures, quickly attaining 
the nuclear equilibrium with $^{17}$O.
The net yield of $^{18}$O is negative for all masses, with a trend
mirroring that of $^{17}$O.
Moreover, as in the case of $^{15}$N, the depletion of $^{18}$O is
more pronounced at higher metallicities due to the more efficient
CNO cycling.

\end{itemize}

\subsection{Secondary and primary components}
\label{primy}
As far as the CNO nuclei are concerned, 
we can distinguish for each element $k$
the secondary, $M_{\rm y}^{\rm S}(k)$, and primary,  
$M_{\rm y}^{\rm P}(k)$, components
of the stellar yield 
\begin{equation}
M_{\rm y}(k) = M_{\rm y}^{\rm S}(k) + M_{\rm y}^{\rm P}(k)
\end{equation}
calculated with (see Eq.~(\ref{defy})) 
\begin{equation}
M_{\rm y}^{\rm S}(k) = \int_{0}^{\tau(M_{\rm i})} 
[X^{\rm S}(k) - X^{\rm S,0}(k)] \,\, 
\frac{{\rm d}M}{{\rm d}t} {\rm d}t
\end{equation}
and 
\begin{equation}
M_{\rm y}^{\rm P}(k) = \int_{0}^{\tau(M_{\rm i})} 
[X^{\rm P}(k) - X^{\rm P,0}(k)] \,\,
\frac{{\rm d} M}{{\rm d}t} {\rm d}t
\end{equation}
where $X^{\rm S}(k)$ and $X^{\rm P}(k)$ denote the 
secondary and primary current 
surface abundances, respectively.
Moreover, we have $X^{\rm S, 0}(k) = X^{\rm 0}(k)$ and 
$X^{\rm P, 0}(k) = 0$, as follows from the definitions of secondary and primary
abundances.

We notice that  $M_{\rm y}^{\rm P}(k)$ can be only $\ge 0$, 
whereas $M_{\rm y}^{\rm S}(k)$
can be either $\ge 0$ or $< 0$, in the 
respective cases that the mass-averaged secondary abundance 
of the element in the ejecta is greater, equal
or smaller than its original value.

A few basic remarks should be made at this point.
Both the first and second
dredge-up  affect (by increasing or decreasing)  only the
secondary components of the CNO surface abundances.
In fact, in these episodes the
envelope is polluted by material which has undergone CNO-cycling, with
a net change in the relative abundances of the CNO 
isotopes  synthesized from metal seeds originally present in the star.
On the contrary, the third dredge-up enriches the
chemical composition of the envelope with  $^{12}$C and $^{16}$O
of primary origin (synthesized by $\alpha$-capture reactions).
Finally, HBB affects the abundance distribution of the CNO
isotopes of both secondary and primary synthesis.

Keeping in mind these concepts, 
it turns out that: 
\begin{itemize}
\item stars with initial masses $M_{\rm i} < M_{\rm dred}^{\rm min}$
can produce CNO yields of secondary origin only; \\
\item for stars with initial masses 
$M_{\rm dred}^{\rm min} \la M_{\rm i} \la M_{\rm HBB}^{\rm min}$  
the yields of $^{12}$C and $^{16}$O should include a primary component,
as a consequence of
the third dredge-up during the TP-AGB phase; and \\ 
\item for stars with initial masses
$M_{\rm i} \ga M_{\rm HBB}^{\rm min}$ 
undergoing HBB during the TP-AGB phase,
the yields of the CNO isotopes should all display primary components, 
because of the injection of primary $^{12}$C and $^{16}$O nuclei 
(at each dredge-up event) into 
regions where the CNO cycle is operating.
\end{itemize}

\begin{table*}[ht]
\caption{Summary of the main prescriptions adopted in the synthetic
TP-AGB models here considered for comparison.}
\begin{tabular}{|l|c|c|c|}
\hline
\noalign{\smallskip}
PROCESS/QUANTITY & \multicolumn{3}{|c|}{MODEL PRESCRIPTION} \\
\noalign{\smallskip}
\hline
\noalign{\smallskip}
\hline
\noalign{\smallskip}
      &  RV81 & HG97 & M2K (this work)\\
\noalign{\smallskip}
\hline
\noalign{\smallskip}
 $M_{\rm up}$ & 8 $M_{\odot}$ & 7 $M_{\odot}$ & 5 $M_{\odot}$ \\
\noalign{\smallskip}
\hline
\noalign{\smallskip}
$M_{\rm c}-L$ relation, $M_{\rm c}-T_{\rm ip}$ relation
& no metallicity dependence  &  with metallicity dependence & 
  with metallicity dependence \\
\noalign{\smallskip}
\hline
\noalign{\smallskip}
 mass loss & Reimers (1975) $\eta=1/3-2/3$ & Reimers (1975) $\eta = 5 $ & 
                                     Vassiliadis \& Wood (1993) \\
\noalign{\smallskip}
\hline
\noalign{\smallskip}
 3 D.up:  efficiency $\lambda$ & $\sim 0.3 - 0.5$  & 0.75 &  0.65 for $Z=0.004$, any $M$\\
                    & function of $M_{\rm c}$, any $Z$ &  for any $Z$, $M$ & 0.55 for $Z=0.008$, any $M$\\
\noalign{\smallskip}
\hline
\noalign{\smallskip}
 3 D.up:  minimum core mass $M_{\rm c}^{\rm min}$ & 0.60 $M_{\odot}$ & 0.58 $M_{\odot}$ & from envelope integrations \\
                                & for any $Z$, $M$ & for any $Z$,$M$  & function of $M$ and $Z$ \\
\noalign{\smallskip}
\hline
\noalign{\smallskip}
HBB: overluminosity & no & no & yes  \\
\noalign{\smallskip}
\hline
\noalign{\smallskip}
HBB: nucleosynthesis & nuclear network & parameterised approx. & nuclear network\\
\noalign{\smallskip}
\hline
\end{tabular}
\label{tab-mod}
\end{table*}

Comparing the secondary and primary  components 
of the CNO yields four cases can be met (Tables A1 -- A12;  
see also Marigo 1998b):
\begin{enumerate}
\item $M_{\rm y}^{\rm S} \neq 0$ {\rm and} $M_{\rm y}^{\rm P} = 0$
so that $M_{\rm y} = M_{\rm y}^{\rm S}$
\item $M_{\rm y}^{\rm S} > 0$ {\rm and} $M_{\rm y}^{\rm P} > 0$
so that $M_{\rm y} > 0$
\item $M_{\rm y}^{\rm S} < 0$ {\rm and} $M_{\rm y}^{\rm P} > 0$
so that $M_{\rm y} > 0$
\item $M_{\rm y}^{\rm S} < 0$ {\rm and} $M_{\rm y}^{\rm P} > 0$
so that $M_{\rm y} < 0$
\end{enumerate}

The first (1.) case applies to low-mass stars with 
$M_{\rm i} < M_{\rm dred}^{\rm min}$, i.e. never experiencing 
the third dredge-up during the TP-AGB phase. No primary component
of stellar yields is expected.  

The second (2.) case corresponds 
to both  primary and secondary production.
In stars with $M_{\rm i} > M_{\rm HBB}^{\rm min}$ it applies, for instance, 
to $^{13}$C, $^{17}$O, and $^{14}$N.
In general, for these elements a positive secondary contribution 
may be provided by the first
(and possibly second) dredge-up and HBB, the latter process
being also responsible for the primary synthesis of these elements
(starting from primary $^{12}$C and $^{16}$O injected by the third dredge-up).

As far as $^{13}$C is concerned (see also Sect.~\ref{stelyie}) we notice 
that a suitable interplay
between the strength of HBB and mass loss can occasionally
result in very favourable conditions for the production of $^{13}$C,
 giving a peak of the related yields (for instance, at 
the model $4 M_{\odot}$ in the case $Z=0.019, \alpha=2.0$).

Concerning the yields of $^{14}$N, 
it should be remarked that the contribution 
from intermediate-mass stars may be relevant 
in view of interpreting, with the aid of
chemical evolutionary models of galaxies, the observed trend  
in the $\log$(N/O) vs. $\log$(O/H) diagram (see, for instance,
Vila-Costas \& Edmunds 1993; Henry et al. 2000b), where 
the large scatter of data points towards lower metallicities
would imply the existence of a significant 
primary component in the measured nitrogen abundances.

The third (3.) possibility  corresponds to a dominant primary production.
In stars with initial masses 
$M_{\rm dred}^{\rm min} \la M_{\rm i} \la M_{\rm HBB}^{\rm min}$,
this case applies to $^{12}$C surface abundance, which is first decreased
by the negative secondary contribution from the first and second dredge-up, 
and subsequently increased by the third dredge-up injecting 
primary nuclei into the envelope.
A similar situation occurs for the yield $^{16}$O in the 
same range of stellar masses, as 
the effect of the third dredge-up prevails over that
of the previous mixing episodes (first and second). 
 
Finally, the fourth (4.) case corresponds to a  dominant secondary depletion.
This refers to  $^{15}$N, $^{18}$O for stars of all masses, and to
$^{16}$O for stars with $M_{\rm i} \ga M_{\rm HBB}^{\rm min}$ if
the reduction of the original abundance 
caused by the first and second dredge-up dominates over  
the injection of primary oxygen via the third dredge-up 
(even possibly partially destroyed by HBB). 
Under these circumstances, 
no enrichment of the interstellar
medium is expected for these elemental species.
 
Tables A10 --  A12 give 
the net yield for each element of the CNO group (T entry),
together with the secondary (S entry) and primary (P entry)
components, for stars with initial masses 
$3.5 M_{\odot} \le M_{\rm i} \le 5 M_{\odot}$, for various values
of the original metallicity and  mixing length parameter. 

\section{Comparison with other calculations}
\label{compare_yields}
We will compare the stellar yields presented in this work 
(hereinafter also M2K)   
with those available in two widely used studies, namely the pioneer 
work by Renzini \& Voli (1981; RV81), and the more recent one by van de Hoek
\& Groenewegen (1997; HG97). 
Before making a direct comparison between 
the yields of various elemental species, we consider it useful first
to recall the relevant prescriptions adopted in the mentioned AGB models, 
and consequently analyse their effects 
by showing how the predictions of different models compare with 
basic observables. 
 
\subsection{Model prescriptions}
\label{modpre}
Table \ref{tab-mod} summarises the relevant assumptions adopted in the AGB 
calculations, which the three different sets of stellar yields under
consideration are derived from. 
For further details the reader should refer
to the original papers and references therein.

\subsubsection{The limiting mass $M_{\rm up}$}
\label{sssect_mup}
First of all, let us consider the quantity $M_{\rm up}$
(see also Sect.~\ref{somedef}), that corresponds to  
the maximum initial mass of a star to develop a degenerate C-O core,
hence experience the AGB phase.
This critical mass heavily depends on the previous evolutionary
history, mainly the extension of the convective core during 
the H-burning phase. Evolutionary models (classical models)
that adopt the Schwarzschild
criterion to define the convective boundaries 
(e.g. Dominguez et al. 1999)
predict higher values for $M_{\rm up}$ than those models that assume 
some overshoot beyond the last formally stable layer against convection
(e.g. Girardi et al. 2000).

Both RV81 and HG97 yields are based on classical models 
and have $M_{\rm up} \sim 7-8 M_{\odot}$, whereas the pre-AGB evolutionary  
models used in this work (M2K) adopt a convective overshoot 
scheme so that $M_{\rm up } \sim 5 M_{\odot}$.
We remark that 
models with masses up to $M_{\rm up}$ cover, 
by definition, 
the whole class of low- and intermediate-mass stars. In other words, 
models with $M_{\rm i} > M_{\rm up}$ 
(for whatever predicted $M_{\rm up}$) would eventually meet the fate of 
supernova explosion.

Finally, 
a cautionary warning should be made in the context 
of practical application of chemical yields.
When stellar yields from stars of different initial masses
are to be included in galactic models of chemical evolution, attention
should be paid to correctly match sets of yields of different mass intervals
(i.e. low, intermediate, high).
If the models do not belong to a homogeneous grid of stellar calculations,
one should at least care to combine stellar yields
of different origin (i.e. stellar code) 
but with the same predicted value for $M_{\rm up}$.
Otherwise, the relative weight of stars belonging to different 
classes (i.e. with different nucleosynthetic histories) to the
integrated chemical enrichment may be substantially mistaken
(for instance, by over- or under- estimating the role of supernovae).

\subsubsection{Analytical relations}
Fundamental relations in synthetic AGB models are the core mass - luminosity
($M_{\rm c}-L$), and core mass - interpulse period ($M_{\rm c}-T_{\rm ip}$) 
relations. Predictions of stellar yields are significantly influenced by 
these input prescriptions, i.e. the luminosity affects 
the mass-loss rates on the TP-AGB, and interpulse-periods determine the 
temporal recurrence of the third dredge-up. 
 
\begin{figure}
\resizebox{\hsize}{!}{\includegraphics{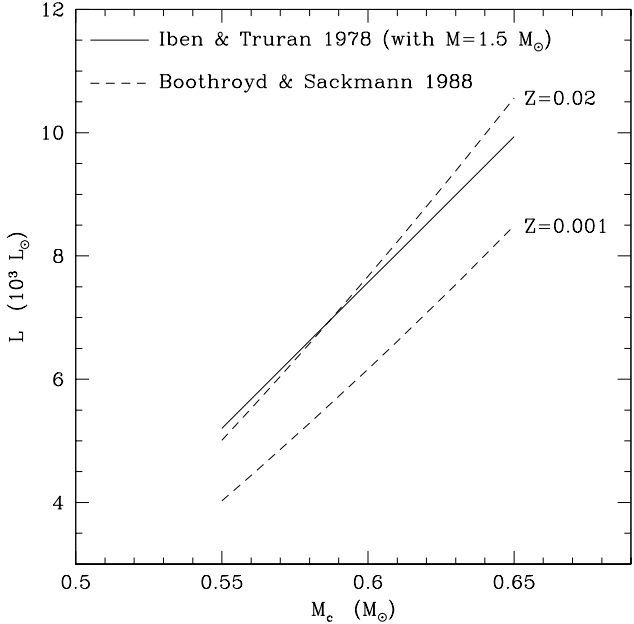}}
\caption{Theoretical core mass - luminosity relations.
The well-known Iben \& Truran (1988) relation is extrapolated for small
core masses and assuming a total mass of 1.5 $M_{\odot}$.
The predicted effect of the chemical composition of the envelope on the 
luminosity is shown according to Boothroyd \& Sackmann (1988a).}
\label{lmc}
\end{figure}

With respect to RV81 prescriptions 
(Iben \& Truran 1978 (IT78), see Fig.~\ref{lmc}), 
M2K and HG97 models are based on  more recent relations  
(Boothroyd \& Sackmann 1988a; Wagenhuber \& Groenewegen 1998) 
in which a notable improvement is the inclusion of a composition dependence. 
For a given core mass, 
the quiescent luminosity / interpulse period of a TP-AGB star is found to
increase /  decrease  at increasing  metallicity.
According to evolutionary calculations 
by Boothroyd \& Sackmann (1988a), for instance,  
the quiescent TP-AGB luminosity for $Z = 0.02$ is about 20 $\%$ 
higher than for $Z = 0.001$ (see Fig.~\ref{lmc}). 

%
%
\subsubsection{Mass loss on the AGB}
The adopted prescription for mass loss on the AGB crucially
influences the predictions of stellar yields, as it 
affects the total number of thermal pulses (hence dredge-up episodes) 
suffered by a TP-AGB star, hence
the growth of its core mass and AGB lifetime.

In this work we adopt the semi-empirical formalism presented
by Vassiliadis \& Wood (1993) who couple results of 
pulsation theory with observations of variable
AGB stars. In RV81 the classical Reimers' law (1975) is assumed
with the efficiency parameter $\eta$ set equal to 1/3 or 2/3. HG97
as well use the Reimers' law, but with $\eta = 5$, which they
find as the best value
to fulfil basic observational constraints
(see Sects.~\ref{wdmass} and \ref{sssec_minmfin}).
%
\begin{figure}
\resizebox{\hsize }{!}{\includegraphics{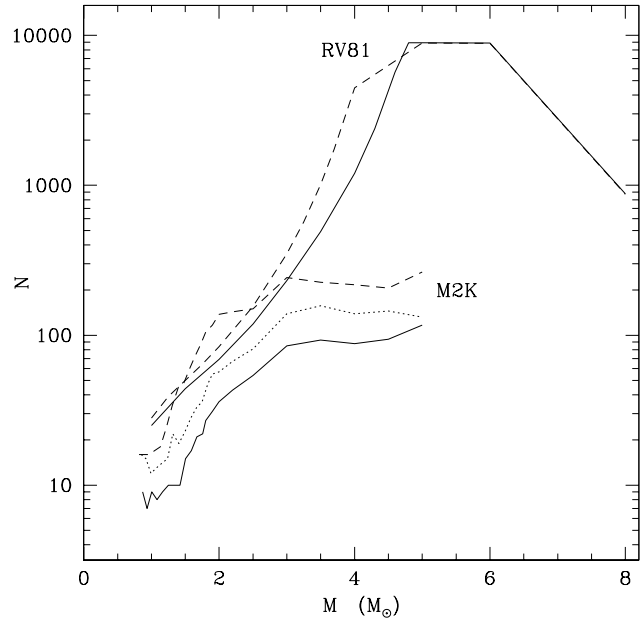}}
\caption{Expected number of thermal pulses (and inter-pulse periods)
as a function of the initial stellar mass for different values
of the initial metallicity (see legenda in Fig.~\protect\ref{fig-time}). 
Results of the present work (M2K) are compared with those of 
Renzini \& Voli (1981; RV81).}
\label{fig-npul}
\end{figure}
%
\begin{figure}
\resizebox{\hsize}{!}{\includegraphics{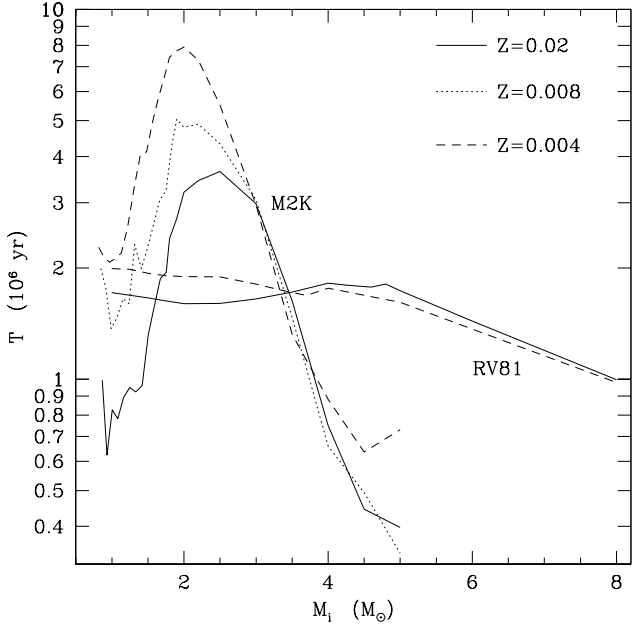}}
\caption{Predicted TP-AGB lifetimes as a function of the initial
stellar mass and metallicity according to the present work (M2K) and
that of Renzini \& Voli (1981; RV81) with the mass-loss parameter $\eta=1/3$.}
\label{fig-time}
\end{figure}
%

To this regard the following remark should be made.
As a matter of fact, the Reimers' prescription was originally designed
to describe mass loss suffered by low-mass stars climbing 
up the RGB, and it is usually calibrated 
in view of reproducing observations
of stars in the subsequent horizontal branch 
phase of quiescent core He-burning
(see Sect.~\ref{models}).

However, as already shown by RV81, the straightforward extension 
of the Reimers' formula to the AGB evolution 
does not suit important constraints.
In fact, the Reimers' prescription with $\eta \la 1$ 
cannot produce the ``super-wind'' 
mass-loss rates ($\dot M \sim 10^{-4}\, -\, 10^{-5}$ $M_{\odot}$ yr$^{-1}$)
measured in stars close to the AGB-tip luminosities, and consequently
it  cannot account for the typical values of masses and radii of 
planetary nebulae at the observed luminosities.

These difficulties have been overcome by later mass-loss prescriptions -- 
specifically designed for AGB stars  
(e.g. Bowen 1988; Fleisher et al. 1992;  Vassiliadis \& Wood 1993;
Bl\"ocker 1995) --  which are all characterised by a more rapid increase
of mass-loss rates during the AGB evolution, then naturally leading to the 
development of the superwind regime.  

The effect of different laws for mass loss is significant with respect
to the expected number of thermal pulses experienced on the TP-AGB
evolution.
As an example (see Fig.~\ref{fig-npul}), 
we can notice that for  a ($5 M_{\odot}, Z=0.02$) model
our calculations yield $N_{\rm p} = 117 - 153 $ (depending on $\alpha$), 
whereas RV81 predict $N_{\rm p} = 8941$ (1631) with the efficiency parameter 
$\eta=1/3$ ($\eta =2/3$).
In general, it turns out that the largest differences in the number of thermal
pulses show up for models with higher stellar masses 
(i.e. $M_{\rm i} \ga 2.5 M_{\odot}$),
that are expected to experience the super-wind regime according to the  
the Vassiliadis \& Wood's prescription.

Moreover, significantly different results 
are obtained by RV81 and M2K  as far as the 
TP-AGB duration is concerned (see Fig.~\ref{fig-time}).
First of all, we can notice that our TP-AGB lifetimes  
present a pronounced trend with the stellar mass, showing a maximum at 
around 2 -- 2.5 $M_{\odot}$ (depending on metallicity). In particular, 
a drastic drop of the TP-AGB
duration is expected in the highest mass domain, i.e. for stars
with HBB.
In RV81 the mass-dependence is much less marked, and  
such a strong reduction of the TP-AGB lifetimes of the 
most massive models is not predicted. This latter point is relevant 
for the interpretation of the high-luminosity wing of 
the observed carbon star luminosity functions (see Sect.~\ref{cslf}).  

We also notice that with the Vassiliadis \& Wood's formalism both 
the total number of the thermal pulses and the TP-AGB lifetimes 
are notably sensitive to the metallicity, i.e. increase with decreasing $Z$.
This feature reflects consequently on the predicted yields  
from stars with the same initial masses but different initial metallicities. 

Finally, it is worth remarking that  the efficiency of mass loss on the AGB 
crucially affects the masses of the bare C-O cores left at the end
of this phase. It follows that the empirical initial-final mass relation
sets important constraints to the theoretical prescriptions for stellar
winds (see Sect.~\ref{sssec_minmfin}).
It is important also to recall that 
according to RV81 in stars with $M_{\rm i} \ga (5-8) M_{\odot}$ 
the C-O degenerate core grows up to 
the Chandrasekhar limit ($\sim 1.4 M_{\odot}$; see Fig.~\ref{minmfin}), 
leading to explosive carbon ignition (Type I-1/2 Supernova event;
see Iben \& Renzini 1983).
On the contrary, according to M2K and HG97 this circumstance
is always prevented by the earlier removal of the 
whole stellar envelope.

\subsubsection{The treatment of the third dredge-up}
This process is expected to critically affect the actual
yields of many elements, mainly $^{4}$He, $^{12}$C, and $^{16}$O. 
Moreover, the reduction of the core mass caused by the third dredge-up
concurs to determine the final mass of the remnant 
left at the end of the AGB. 

The largest differences in the analytical treatment 
of this process, in the models here considered,  
can be summarised as follows.
In RV81 model
the onset of the third dredge-up in low-mass stars possibly occurs 
later (higher $M_{\rm c}^{\rm min}$)
and with a lower efficiency (lower $\lambda$) than in the HG97 and M2K 
calculations, that are carried out with similar values of the parameters. 
(see Table \ref{tab-mod}).

This can be explained considering the different usage  
of the quantities  $M_{\rm c}^{\rm min}$, and $\lambda$ in the models.
As already mentioned in Sect.~\ref{intro},
in RV81 these parameters  were  derived
according to complete stellar models currently available at that epoch. 
The subsequent comparison between the predictions of synthetic models
and observations pointed out at the 
so-called ``carbon star mystery'', as denominated by Iben (1981),
i.e. too few faint and too many bright carbon stars
expected than observed.

Differently, the later analyses carried out by HG97 and M2K 
move from another perspective,
that is to consider   
$M_{\rm c}^{\rm min}$ and $\lambda$ as free parameters which  
should be calibrated in order to reproduce observations 
of carbon stars. The aim is to provide  
indications on the average characteristics of the third dredge-up, 
so as to remove the theoretical discrepancy related
to the ``carbon star mystery'' (see Sect.~\ref{cslf}).

\subsubsection{The treatment of hot bottom burning}
The most notable effect of this process on stellar yields
involves $^{14}$N and $^{4}$He, which are newly synthesised at the expense
of hydrogen and, in general, of the other CNO catalysts.  
It is important to stress that
the occurrence of HBB in the most massive AGB stars ($M > 3.5 - 4.5 M_{\odot}$)
has not only a direct effect on stellar yields -- 
via changing  the chemical abundances in the envelope -- 
but also determines  an indirect action,  
affecting the energetics of these stars, hence their evolutionary
properties.

\begin{figure}
\resizebox{\hsize}{!}{\includegraphics{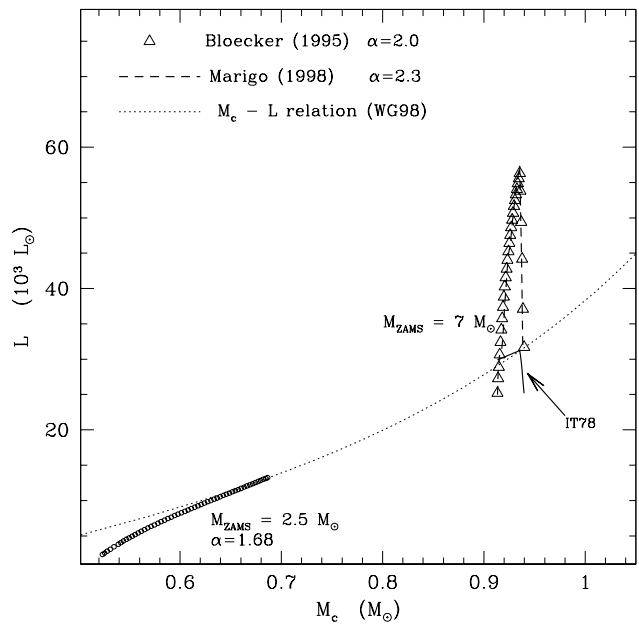}}
\caption{Quiescent luminosity as a function of 
the core mass during the  TP-AGB phase.
The predictions for the 7 $M_{\odot}, Z=0.021$ model with HBB
according to full evolutionary calculations by Bl\"ocker (1995) are compared
to the those of synthetic calculations. 
The luminosity evolution for a $2.5 M_{\odot}$ model 
is also shown (Marigo et al 1999a). 
The reference $M_{\rm c}-L$ relation is taken 
by Wagenhuber \& Groenewegen (1998). 
The adopted mass-loss prescription is that by Baud \& Habing (1983).
See text for further explanation.}
\label{hbb7}
\end{figure}

In fact, as a consequence of HBB, the $M_{\rm c}-L$ relation 
breaks down, a feature clearly shown by complete AGB stellar calculations
(e.g. Bl\"ocker \& Sc\"onberner 1991; see also Fig.~\ref{hbb7}). 
In these massive AGB models the luminosity evolution is characterised 
by a steeper increase 
with the core mass (above the $M_{\rm c}-L$ relation) up to a maximum,
followed by a decline as soon as the envelope mass 
is significantly reduced by mass loss. Eventually the $M_{\rm c}-L$ relation
is recovered (e.g. Vassiliadis \& Wood 1993; Marigo 1998b).
This behaviour is exemplified in 
Fig.~\ref{hbb7} where we report the results  
of complete evolutionary
calculations performed by Bl\"ocker (1995) for a $7 M_{\odot}$
model with solar metallicity (triangles). 

Actually, the effect of the predicted luminosity evolution 
on stellar yields is at least 
two-fold. In fact, the stellar luminosity is closely related
to the temperature at the base of the envelope, which the nuclear reaction 
rates crucially depend on.
Moreover, the overluminosity of AGB stars with HBB can
trigger high mass-loss rates, thus favouring the onset of the super-wind
regime with consequent reduction of the TP-AGB lifetimes.

Such overluminosity  effect caused by HBB 
is included neither in RV81 nor in GdJ93, where
the luminosity evolution  
is assumed to follow the $M_{\rm c}-L$ relation by IT78  
(with some revision for the composition dependence in the GdJ93 work).
In Fig.~\ref{hbb7} we show the behaviour 
of the luminosity for the $7 M_{\odot}$ model (solid line)
as it would result adopting the IT78 relation with the 
the same values for current $M_{\rm c}$ and $M$ as in the Bl\"ocker (1995)
model sequence. The discrepancy is notable.
To overcome this limitation of synthetic models
Marigo et al. (1998; see also Marigo 1998ab) developed a solution 
scheme based on envelope integrations, so that the overluminosity 
produced by HBB is taken into account and the results of complete
stellar calculations are recovered (dashed line in Fig.~\ref{hbb7}).

\subsection{Observational constraints}
In the following we will examine the AGB synthetic models
under consideration (RV81, HG97,  M2K (this work)) in relation
to their capability of reproducing important observational constraints, which
are closely related to the stellar yields.

\begin{figure}
\resizebox{\hsize}{!}{\includegraphics{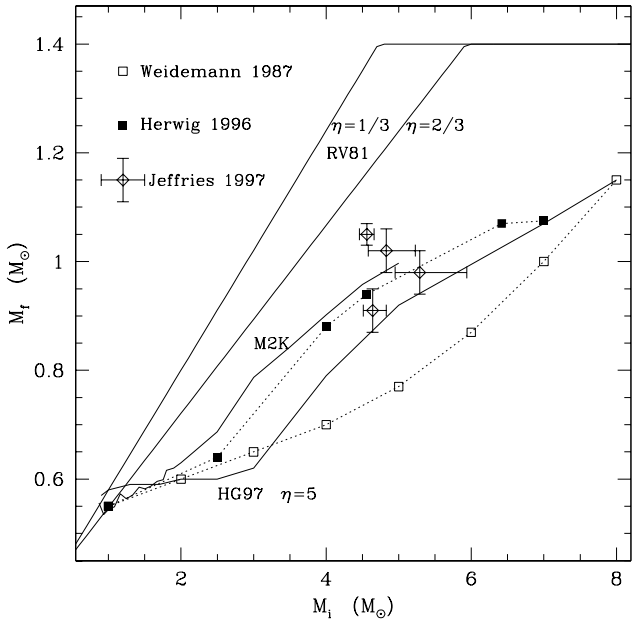}}
\caption{Initial-final mass relation for low- and intermediate-mass
stars with solar metallicity. Semi-empirical calibrations for the 
solar neighbourhood are taken from Weidemann (1987), Herwig (1996),
and Jeffries (1997). Solid lines refer to theoretical predictions.
When the Reimers' prescription for mass-loss is adopted, the corresponding
efficiency parameter $\eta$ is indicated. See text for more details.}
\label{minmfin}
\end{figure}

\subsubsection{The initial - final mass relation}
\label{sssec_minmfin}
\begin{figure*}
\begin{minipage}{0.67\textwidth}
\resizebox{\hsize}{!}{\includegraphics{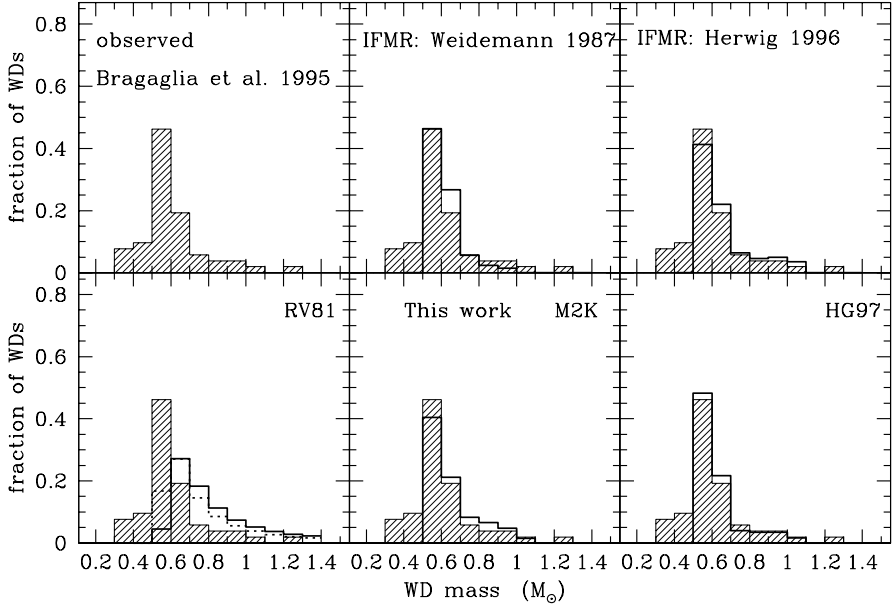}}
\end{minipage}
\hfil
\begin{minipage}{0.30\textwidth}
\caption{WD mass distribution for solar metallicity. 
The observed data in the 
solar neighbourhood are taken from the sample of bright DA WDs 
by Bragaglia et al (1995). 
The shaded area beneath the observed histogram is set equal to unity.
The middle and right panels on the top show the WD mass distributions
(solid line)
derived by adopting the semi-empirical IFMRs by Weidemann (1987) and
Herwig (1996), respectively.
The bottom panels display    
the distributions obtained by assuming the theoretical IFMRs
from the quoted works. All the solid line histograms   
are normalised to the 
fraction of observed WDs with masses larger than 0.5 $M_{\odot}$.
See text for more details.}
\label{wdmd}
\end{minipage}
\end{figure*}

The initial-final mass relation (IFMR) 
of low- and intermediate-mass stars is  intimately linked to the 
chemical yields, as it determines the total amount of matter 
ejected by a star during its entire evolution. In a complementary way, it 
gives information on the reservoir of stellar remnants, irreversibly
lost by the star-forming gas. Moreover, 
assessing the upper mass limit for WD progenitors ($M_{\rm WD}$)
is an important point, since it 
affects the expected rate of type II supernovae.
All these aspects are fundamental issues for chemical evolutionary
models.   

Figure \ref{minmfin} shows a few  empirical calibrations
of the IFMR for the solar neighbourhood.
The first striking point is that the more recent determinations
significantly differ from the earlier work by Weidemann (1987).
For instance, the revised relation by Herwig (1996) presents 
a flatter slope up to Hyades location 
($M_{\rm i} \sim 3 M_{\odot}$, $M_{\rm f} \sim 0.7 M_{\odot}$),
followed by a steeper rise, and a final flattening
towards higher initial masses ($M_{\rm i} > 4 M_{\odot}$).
The presence of an inflection point at the Hyades mean location seems
to be confirmed also by Reid (1996), as discussed by Weidemann (1997).
 
The second point to be made deals with
the critical mass $M_{\rm WD}$\footnote{For the sake fo simplicity, 
we limit our discussion to 
$M_{\rm WD}$ for carbon/oxygen white dwarfs, i.e. not considering
the neon/oxygen white dwarfs which would derive from ``super-AGB''
stars according to Garcia-Berro et al. (1997) evolutionary calculations.}, 
that is the maximum initial mass of WD progenitors. 
At present, this limiting value is still rather uncertain
(most likely in the range 5 -- 8 $M_{\odot}$), since it heavily
depends on model details.
In particular, as already discussed by Weidemann (1987), 
the definition of convective boundaries --  
via either the Schwarzschild criterion or an overshooting scheme --
plays a crucial role. It turns out that
with the latter choice $M_{\rm WD}$ is lower than assuming the former 
classical assumption.
However, other parameters may affect the predictions for $M_{\rm WD}$.
For instance, 
the recent metallicity re-determination  (i.e. half-solar) of the  
young open cluster NGC 2516 by Jeffries 
(1997; see Fig.~\ref{minmfin}) has lead to assign it
a younger age. As a consequence, Jeffries (1997) derives   
$M_{\rm WD}$ around 5 -- 6 $M_{\odot}$, that is considerably lower than 
$M_{\rm WD} \sim 7-8 \, M_{\odot}$ as estimated in previous studies
(e.g. Weidemann 1987; Koester \& Reimers 1996).
However, it should be noticed that in a more recent 
paper Jeffries et al. (1998) still do not exclude 
that the metallicity of NGC 2516 might be nearly solar.

Figure \ref{minmfin} displays the theoretical IMFRs 
as derived by RV81, HG97, and M2K 
for low- and intermediate mass-models with initial solar metallicity.
We can notice that the both HG97 and M2K are satisfactorily consistent 
with the trend of the most recent observational relations, 
whereas RV81 is far from reproducing the empirical
data. In particular, the RV81 relation shows a quick divergency of the
final mass at increasing initial mass, with the most massive stars being  
able to build C-O cores up to the Chandrasekhar limit of 1.4 $M_{\odot}$. 
The final fate of these stars would correspond to the occurrence of
type I-1/2 supernova events, which seems not to be supported
by the observations. 

\subsubsection{The white dwarf mass distribution}
\label{wdmass}
The white dwarf mass distribution (WDMD) is closely related to the IFMR
from which it can be theoretically derived, 
provided that assumptions on the initial
mass function (IMF) and star formation rate (SFR) are made to perform
the integration over mass and time.
The observed WDMD in the solar neighbourhood 
(Bergeron et al. 1992; see Fig.~\ref{wdmd}) is narrowly peaked in the
mass range 0.5 -- 0.6 $M_{\odot}$ (adopting a mass bin
of 0.1 $M_{\odot}$), which contains more than 45 $\%$ of the 
observed objects. 

We notice that, with a finer bin sampling (i.e. 0.05 $M_{\odot}$), the 
location of observed peak would fall between 0.5 and 0.55 $M_{\odot}$,
which may be difficult to be theoretically reproduced.
In fact, 
according to stellar evolutionary models, these values would be consistent 
with the minimum remnant mass produced by progenitor stars 
as old as the age of the Galaxy ($\sim 15$ Gyr corresponding to 
initial masses $\sim 0.9 \, M_{\odot}$), provided that 
their C-O cores do not grow in mass during the TP-AGB phase.
In fact, these stars are predicted to enter the TP-AGB phase 
with a core mass of already $\sim 0.52 \, M_{\odot}$ (Girardi et al. 2000).
Then, the location of the observed peak in the range 0.5 -- 0.55 $M_{\odot}$
might be explained by theory 
only if assuming i) that mass loss suffered by AGB low-mass stars  
is so strong that they leave this phase as soon as they enter it, 
or ii) very efficient dredge-up prevents the growth in mass of the core
(Herwig et al. 1997).
On the other hand, as suggested by Bragaglia et al. (1995),
the origin of such discrepancy
between theory and observations would be most likely due to 
a systematic underestimation of the surface gravities derived from WD models.

Given this point of uncertainty and considering that WDs 
with $M < 0.4 M_{\odot}$
are probably helium WDs derived from binary evolution, 
in this work both theoretical
and observed WDMDs (see Fig.~\ref{wdmd}) are derived adopting a mass bin
of 0.1 $M_{\odot}$, and normalising them 
to the observed fraction of WDs with $M \ge 0.5 M_{\odot}$.

\begin{figure}
\resizebox{\hsize}{!}{\includegraphics{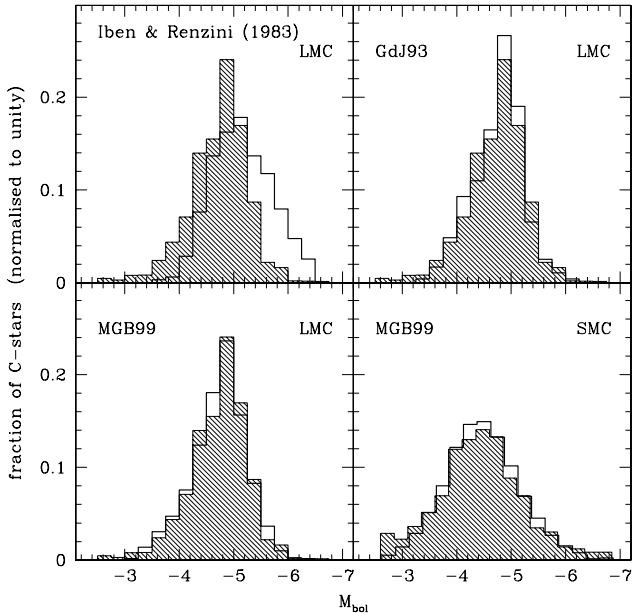}}
\caption{Luminosity functions of field carbon stars 
in the Magellanic Clouds. The observed data 
(shaded histograms) are taken from Costa \& Frogel (1996) for the LMC, 
and Reiberot (1993) for the SMC. Theoretical distributions 
(thick solid line) are shown for comparison.
Top left-hand side panel:
Iben \& Renzini (1983) calculations for the LMC -- based on a synthetic  
AGB model very similar to that of Renzini \& Voli (1981; RV81) -- 
with the parameter set
($\eta =2/3$, $\alpha =2$, $\epsilon = 0.1$).
Top right-hand side panel: Groenewegen \& de Jong (1993; GdJ93) best
fitting distribution for the LMC carbon stars. 
Bottom panels: Marigo et al. (1999a) 
best fits to the CSLFs in the LMC and SMC.
See text for further explanation.}
\label{fig_cslf}
\end{figure}

The predicted fraction of WDs contained in the $k^{\rm th}$ mass 
bin (from $M_{\rm f}^{\rm k}$ to $M_{\rm f}^{\rm k+1}$) is calculated with:
\begin{equation}
 N^{k} \propto\frac{(M_{\rm i}^{k+1}- M_{\rm i}^{k})} 
{(M_{\rm f}^{k+1}- M_{\rm f}^{k})} \:\: \phi(M_{\rm i}^{k/2}) 
 \:\: \psi(T_{\rm G} - \tau_{k/2}) \:\: \Delta T_{k/2}  
\label{eq_wdmd}  
\end{equation}  
where ($M_{\rm i}^{k}$,  $M_{\rm i}^{k+1}$) is the corresponding
interval of initial stellar mass; $\phi(M_{\rm i}^{k/2})$ is the
IMF (by number) evaluated at the mean initial mass
$0.5\,\,(M_{\rm i}^{k+1}+ M_{\rm i}^{k})$; $T_{\rm G}$ and 
$\tau_{k/2}$ denote the age of the Galaxy, and the lifetime of
a star with mass $M_{\rm i}^{k/2}$, respectively;
$ \psi(T_{\rm G} - \tau_{k/2})$ is the SFR evaluated at the birth epoch
of the star; and $\Delta T_{k/2}$ is the time interval of
detectability of the WD.

For the sake of simplicity, 
in our calculations we adopt a constant SFR, the IMF given by the 
Salpeter's law ($\phi(M) \propto M^{-2.5}$), and  suppose that any 
WD formed in the past is still detectable at the present time.
This implies we assume that the WD fading time is always 
much longer than the WD's age. In this case 
$\Delta T_{\rm k/2} = T_{\rm G} - \tau_{\rm k/2}$, i.e. 
the WD has been detectable since the death of the progenitor.  

Moreover, it is worth noticing the following points. 
Under the assumption of a constant SFR, 
the WDMD essentially depends on i) the slope, 
${\rm d} M_{\rm i}/{\rm d} M_{\rm f}$, of the IFMR,
 ii) the IMF, and iii) the lifetimes of the stellar progenitors 
relative to the age of the Galaxy.
 
The first factor favours the population of those mass bins in which the 
slope of the IFMR is flatter, i.e. where stars with 
different initial masses build up WDs with similar masses. This should be 
one of the dominant effects which gives rise to the observed narrow peak
of the WDMD at around $0.5 - 0.6 \, M_{\odot}$, just where the IFMR is rather
flat (see Fig.~ \ref{minmfin}).

The second and third factors tend to produce opposite effects.
The IMF preferentially weighs the formation of WDs with 
lower masses, hence evolved from originally less massive stars  
if the IFMR is a monotonic function. On the contrary, 
the accumulation factor  ($T_{\rm G} - \tau$) 
favours the contribution of WDs of higher 
masses, evolved from more massive stars, hence with shorter lifetimes. 

In Fig.~\ref{wdmd}  the observed WDMD in the solar neighbourhood 
is compared with the distributions derived according to 
Eq.~(\ref{eq_wdmd}) assuming different 
IFMRs, both empirical and theoretical ones.
The relations by Weidemann (1987) and Herwig (1996), though
being quite different, yields WDMDs both acceptably 
consistent with the observed one.
This can be explained considering that the major differences between the 
two IFMRs show up for $M_{\rm i} > 2.5 - 3 M_{\odot}$, corresponding 
to WDs that do not contribute to the mass peak. For 
$M_{\rm f} \le 0.55 M_{\odot}$ the relations are quite similar, showing 
a rather flat trend.

As far as the purely theoretical WDMDs are concerned 
(bottom panels), it turns out that a satisfactory reproduction 
of the observed data is attained by both HG97 and M2K (this work), whereas
a notable discrepancy affects the predictions by RV81.
As already anticipated in the discussion on the predicted 
IFMRs (Sect.~\ref{sssec_minmfin}), in RV81 there is  
a sizable overproduction of WDs more massive
than $0.6 M_{\odot}$, a feature already pointed out by Bragaglia 
et al. (1995). 
\begin{figure*}
\begin{minipage}{0.49 \textwidth}
 \resizebox{\hsize}{!}{\includegraphics{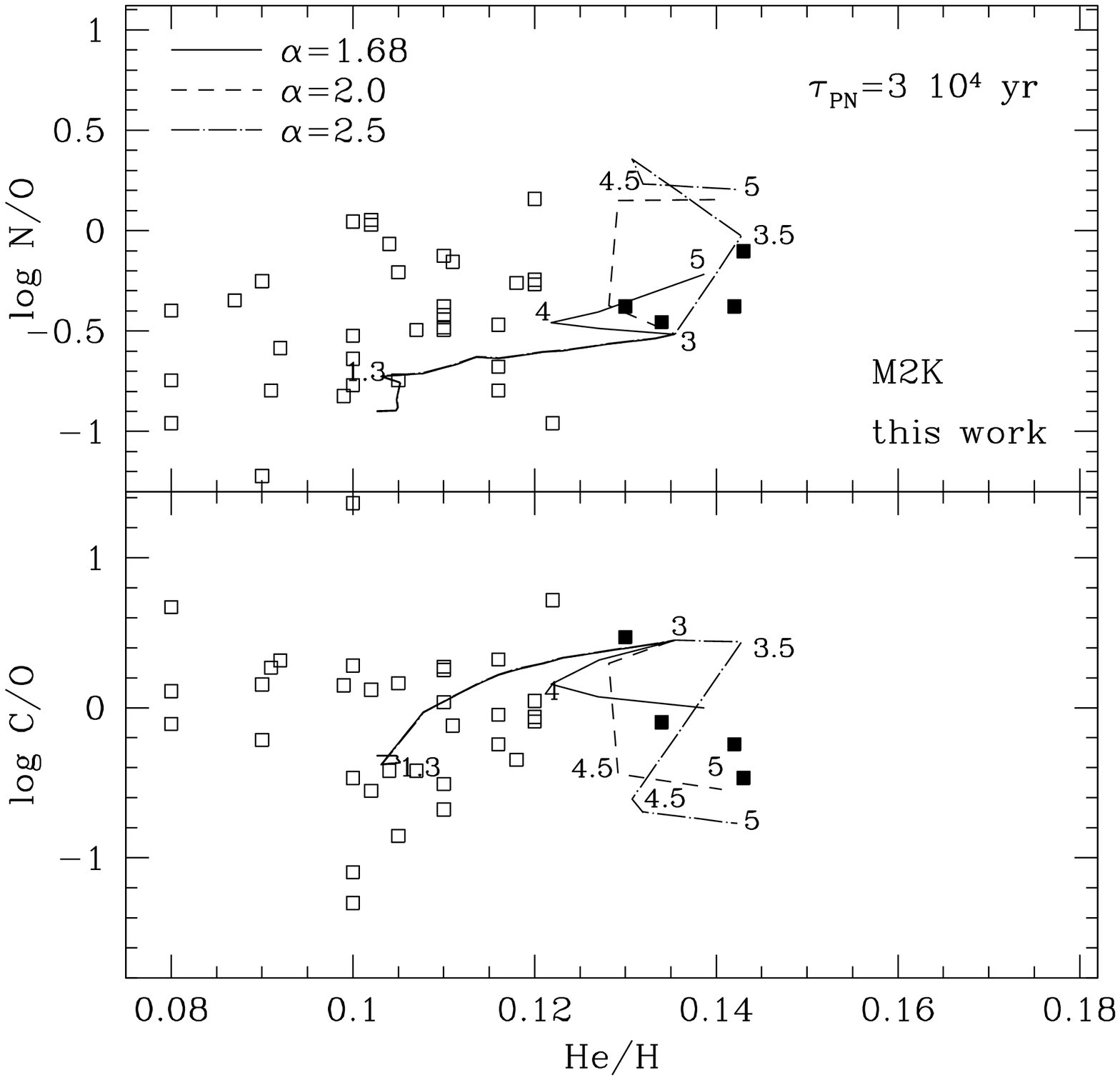}}
\end{minipage}
\hfill
\begin{minipage}{0.49\textwidth}
 \resizebox{\hsize}{!}{\includegraphics{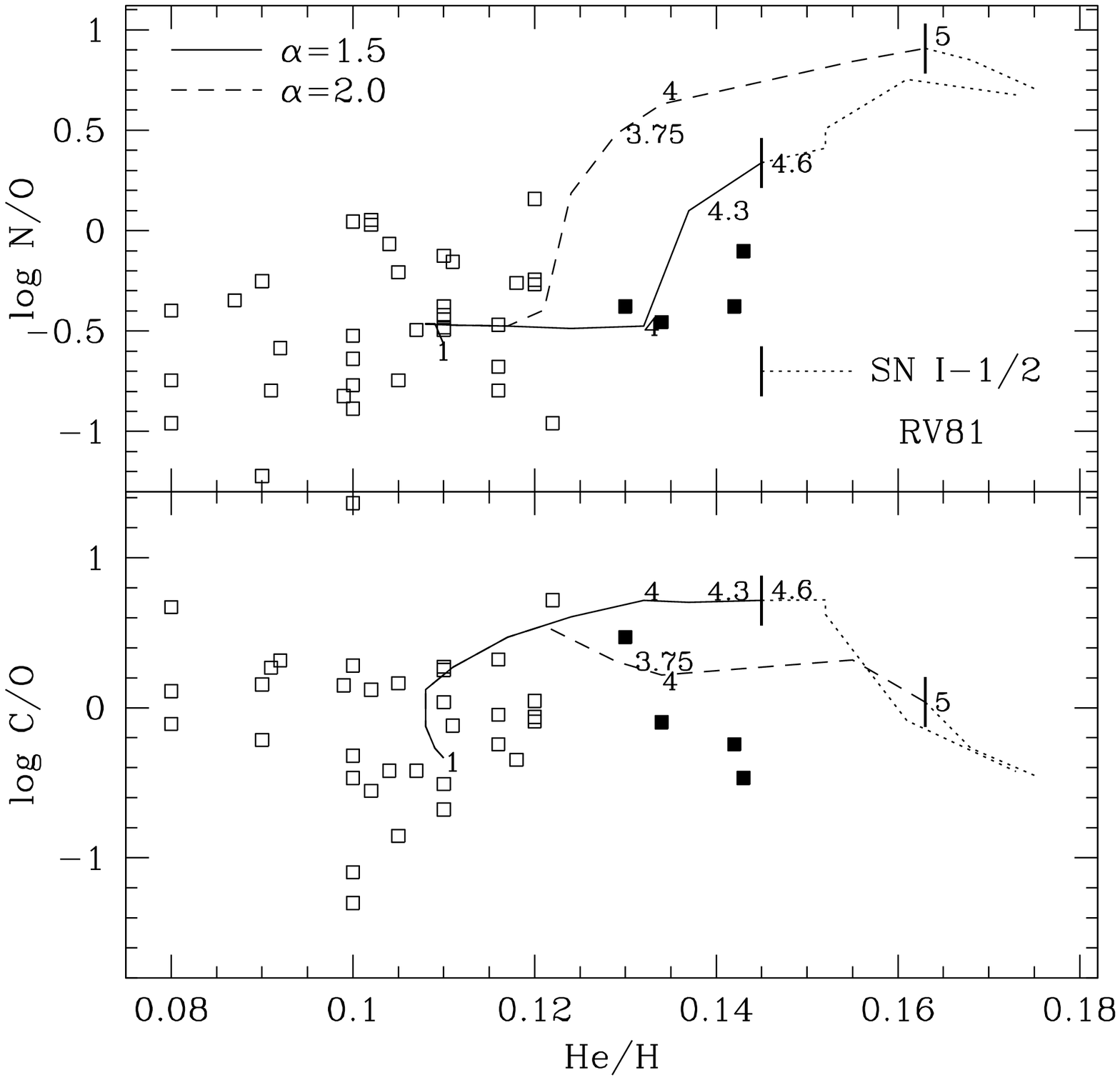}}
\end{minipage}
\caption{Abundance ratios of galactic planetary nebulae.
Observed data (squares) are taken from Kingsburgh \& Barlow (1994), and 
Henry et al. (2000a). Filled squares correspond to PNe with 
$\log({\rm N/O}) > -0.5$ {\it and} ${\rm He/H} > 0.125$.
Predicted PN abundances are shown as a function of the initial mass of
solar-metallicity progenitor stars, as derived in this work
(left-hand side panel), and in Renzini \& Voli (1981, right-hand side panel).
Lines (solid and dashed) connect predicted 
abundances at increasing stellar mass (a few values are indicated nearby) 
for two values of the mixing length parameter. In the case of RV81 model, 
surface abundances just prior Chandrasekhar carbon explosion are also shown 
(dotted line).}
\label{fig-chempn}
\end{figure*}

\subsubsection{The carbon star luminosity function}
\label{cslf}
The carbon star luminosity function (CSLF) is a fundamental observable
as it gives indications on, at least, two basic processes
occurring in TP-AGB stars with different masses, namely: 
the third dredge-up -- that determines the increase of the
surface carbon abundance --,  
and mass loss by stellar winds, that affects the duration, hence
the luminosity excursion during this phase.

Iben (1981) first pointed out the so-called ``carbon star mystery'', 
corresponding to a long-standing
discrepancy between theory and observations, i.e. current stellar models 
predicted a deficit of faint carbon stars,
accompanied by an excess of bright carbon 
(in general AGB) stars.
This situation is exemplified in Fig.~\ref{fig_cslf}, where we report
the predicted CSLF for the LMC, according to the calculations performed
by Iben \& Renzini (1983), with model prescriptions very similar to RV81.
For this particular case, 
the authors adopt the following set of parameters:
efficiency parameter $\eta = 2/3$ in the Reimers (1975) mass-loss
formula; mixing-length parameter, $\alpha=2$; 
minimum core for the third dredge-up to occur,  
$M_{\rm c}^{\rm min} = 0.5 M_{\odot}$; and efficiency of the third dredge-up,
$\lambda$, as a function of the core mass.

The CSLF in the LMC is instead very well fitted (Fig.~\ref{fig_cslf})
by the other two AGB synthetic
models here considered, namely: Groenewegen \& de Jong (1993, GdJ93; 
top right-hand side panel) and
Marigo et al. (1999a, MGB99; bottom left-hand side panel). 
We remind again that, differently from RV81, in these studies 
the third dredge-up is suitably calibrated in order to reproduce the 
CLSF in the LMC. 
Finally, it should be remarked that Marigo et al. (1999a) have extended the
analysis to the CSLF in the SMC (see bottom right-hand
side panel of Fig.~\ref{cslf}), so as to include a metallicity-dependent
treatment of the third dredge-up in their synthetic AGB model.

\subsubsection{The chemical abundances of planetary nebulae}
\label{chempne}
The observed chemical composition of planetary nebulae represents another
crucial constraint for stellar models, as it   
is the record of the nucleosynthetic and mixing processes occurred
during the previous stellar evolution. 

As far as HG97 predictions are concerned, 
a detailed discussion is given in Groenewegen \& de Jong
(1994) and it will not be repeated here.
We restrict here to the results of RV81 and M2K which are 
compared in Fig.~\ref{fig-chempn} with the  
measured  abundances of He, C, and O in galactic planetary nebulae.
Predicted PN abundances can be found in Tables A13 -- A15.   
Since a full analysis is beyond the purpose of this work,
we simply consider the most relevant aspects. 

Both RV81 and M2K results shown here  
are derived from calculations of stellar models with initial solar 
composition. Therefore, they cannot 
reproduce the data points with He/H $\la 0.10-0.11$, since these latter
most likely correspond to progenitor stars with initial subsolar metallicity.

The expected paths of PN abundances
as a function of the initial mass of the progenitor star reflect
the efficiency and duration of the involved processes.
For instance, in both models the C/O ratio first increases at increasing mass
due to the third dredge-up, and then (for $M > 3-4 M_{\odot}$)
it starts to decline because of HBB.

An interesting point is the  anticorrelation between 
the N/O ratio and the C/O ratio exhibited by 
observed PNe with the highest helium content (He/H$ > 0.125$; the so-called
type I PNe according to the classification introduced by Peimbert (1978)).
This trend is reproduced by M2K synthetic calculations, 
tracing the signature of HBB in the most massive AGB stars 
($M \ga 3.5 M_{\odot}$), where carbon in the envelope is quickly converted 
into nitrogen. 

Note that theoretical results are 
notably sensitive to the adopted value for the mixing-length parameter,
i.e. the larger $\alpha$ is, the more efficiently HBB operates, yielding
higher N/O and lower C/O ratios.
Limiting to the observed sample of PNe, we might deduce 
that HBB has operated in the most massive progenitors of
solar metallicity, but with a rather mild efficiency, 
in agreement with the conclusion already mentioned
by Henry et al. (2000a).
In fact, as we can see from Fig.~\ref{fig-chempn}, predictions
for the case ($\alpha = 2.5$) -- which correspond to strong HBB --
lead to N/O ratios quite higher than observed.

Hower, these indications should be considered with some caution,
as the considered sample of PNe might not cover the whole relevant
mass range of the progenitors (see Henry et al. 2000a), and
predictions of PNe abundances are derived under very simple
assumptions (see Appendix A). 
A much better approach will be adopted with the aid of 
a detailed synthetic model of PN evolution, which 
is being developed (Marigo et al. 2001, in preparation; see
Marigo et al. 1999b for a preliminary presentation).
  
Finally, we would remark that RV81 results for the most massive 
stellar models  -- shown in Fig.~\ref{fig-chempn} 
with dotted lines -- {\sl do not} correspond to PN abundances, but rather
to surface abundances just prior the progenitor stars explode as  
SNe I-1/2. Therefore, attention must be paid not to use these data
for a comparison with observed PN abundances.


\subsection{Yields from simple stellar generations}
\label{yield_comp}
We will compare here the stellar yields presented in this work  
with those calculated by RV81 and HG97.
Possible differences are then discussed on the basis of different model
prescriptions and observational constraints, already mentioned
in the previous sections.

We choose not to make a direct comparison 
between yields produced by models with the same initial mass,
because different sets of yields cover different mass-ranges 
in the domain of low- and intermediate-mass stars (see Sect.~\ref{sssect_mup}).
For this reason, it is more meaningful to compare the 
whole chemical contribution provided by low- and intermediate-mass stars 
belonging to a given simple (i.e. coeval) stellar population.
To this aim, we recall that according to
the standard  definition (Tinsley 1980), the {\it yield from
a stellar generation}, $y_{k}$, is the mass
converted into the chemical element $k$ and ejected by all stars per unit mass
locked into stars:
\begin{equation}
y_{k} = (1-R) \frac{ \int_{M_{\rm l}}^{M_{\rm u}} M_{\rm i} 
		p_{k}(M_{\rm i}) \phi(M_{\rm i}) {\rm d} M_{\rm i} }
		   { \int_{M_{\rm l}}^{M_{\rm u}} 
		M_{\rm i} \phi(M_{\rm i}) {\rm d}M_{\rm i} }
\label{ytot}
\end{equation}
In the above equation 
$p_{k}(m)$ is the stellar yield of the $k^{\rm th}$ element 
(see Sect.~\ref{yi}), and
$R$ is the {\it returned fraction}, 
expressing the fraction
of mass that has formed stars and then been ejected:
\begin{equation}
R = \frac{ \int_{M_{\rm l}}^{M_{\rm u}} [M_{\rm i} - W(M_{\rm i})]  
			\phi(M_{\rm i}) {\rm d}M_{\rm i} }
		   { \int_{M_{\rm l}}^{M_{\rm u}} 
			M_{\rm i} \phi(M_{\rm i}) {\rm d}M_{\rm i} }
\end{equation}
where $\phi(M)={\rm d}N/{\rm d}M_{\rm i}$ is the IMF (by number) 
defined between the lower
($M_{\rm l}$) and upper ($M_{\rm u}$) mass limits; $W(M_{\rm i})$ 
is the remnant mass.

In order to weigh the sole contribution from the 
generation of low- and intermediate-mass stars, let us consider
the quantity:
\begin{equation}
y^{\rm lims}_{k} = \frac{ \int_{M_{\rm l}}^{M_{\rm up}} M p_{k}
			(M_{\rm i}) \phi(M_{\rm i}) dM_{\rm i} }
		   { \int_{M_{\rm l}}^{M_{\rm u}} 
		     M_{\rm i} \phi(M_{\rm i}) dM_{\rm i} }
\label{ylims}
\end{equation}
that is similar to Eq.~(\ref{ytot}) with $M_{\rm u} = M_{\rm up}$. 
The adopted integration extremes are  $M_{\rm l}=0.9 M_{\odot}$, $M_{\rm u}
= 100 M_{\odot}$, and $M_{\rm up}$ according to 
the set of stellar yields under consideration 
(see Table \ref{tab-mod} and Sect.~\ref{sssect_mup}). 
The IMF is expressed by the classical Salpeter's law, i.e. 
$\phi(M_{\rm i})={\rm d}N/{\rm d}M_{\rm i} \propto M_{\rm i}^{-(1+x)}$ with $x=1.35$
(Salpeter 1955).

\begin{figure*}
\begin{minipage}{0.67\textwidth}
\resizebox{\hsize}{!}{\includegraphics{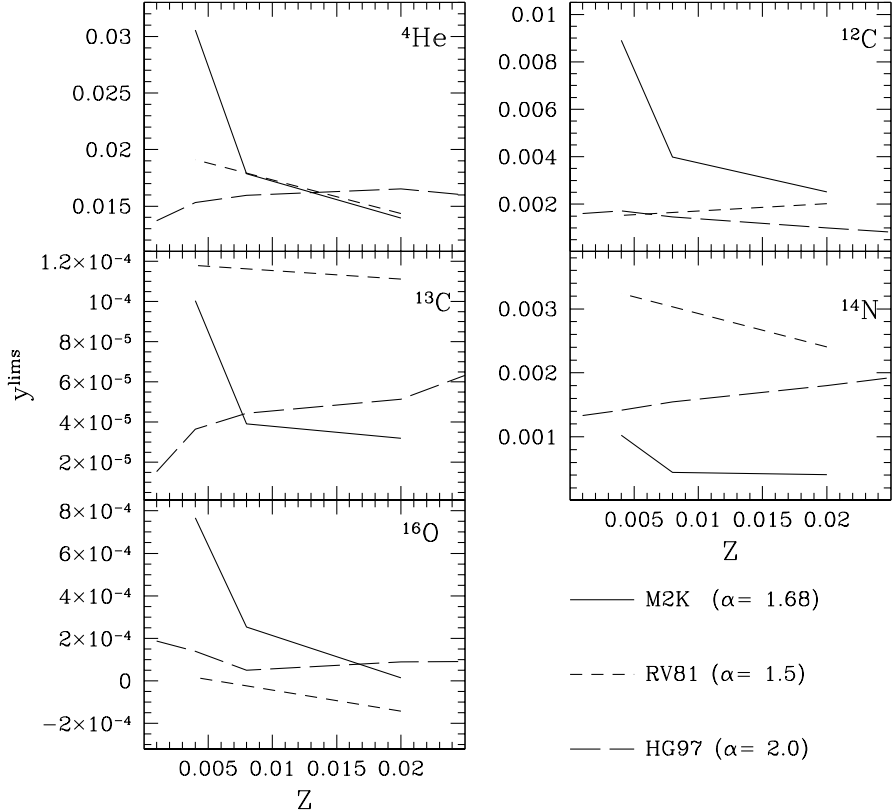}}
\end{minipage}
\hfill
\begin{minipage}{0.30\textwidth}
\caption{\mbox{Integrated yield} contributions from low- and intermediate-mass
stars as a function of the metallicity, as defined by Eq.\protect(\ref{sspy}).
The mixing-length parameters ($\alpha$) adopted by the authors are indicated.}
\label{sspy}
\end{minipage}
\end{figure*}

The quantities $y_{k}^{\rm lims}$ express the relative  
chemical contribution (for a given elemental species $k$) 
from low- and intermediate-mass  stars belonging
to a given simple stellar population.
They are shown in Fig.~\ref{sspy} as a function of the metallicity for
the three sets here considered.

It should be remarked that the
differences between our results (M2K) and those derived by RV81 and HG97
are not only due to the different mass-range covered by low- and intermediate-
mass stars, but mainly reflect substantial differences in the adopted physical
prescriptions as already illustrated in Sect.~\ref{modpre}.

Differences essentially show up both in metallicity trends and 
absolute values of $y^{\rm lims}_{k}$.
Compared to previous calculations, M2K yields show
a pronounced dependence on the metallicity, i.e. positive yields increase with 
decreasing $Z$. Conversely, the RV81 and HG97 sets 
present weak trends with $Z$.

The metallicity dependence can be explained as follows.
On one side, AGB lifetimes of low-mass stars increase
at decreasing  metallicities, as mass-loss rates are expected to be lower.
This fact leads to  a larger number of dredge-up episodes.
Moreover, both the onset and the efficiency of the third dredge-up
are favoured at lower metallicities. These 
factors concur to produce a greater enrichment in carbon.  
On the other side, HBB in more
massive AGB stars becomes more efficient
at lower metallicities, leading to a greater enrichment in nitrogen.
The combination of all factors favours higher positive yields
of helium at lower $Z$.

As far as the single elemental species are concerned, we can notice:
\begin{itemize}
\item M2K yields of $^4$He are larger than those by HG97 and RV81 
towards lower $Z$, likely due to 
the earlier activation and larger efficiency of the third dredge-up
in our models.
With respect to RV81 and HG97 predictions, our yields of $^4$He
present a significant trend with $Z$.
\item M2K  yields of $^{12}$C are systematically higher than those
of RV81 and HG97 because of the 
earlier onset (and average greater efficiency than in RV81) 
of the third dredge-up.
\item The dominant contribution to the yields of $^{14}$N comes
from HBB in the most massive AGB stars.
Differences in the results reflect  different efficiencies
of nuclear reactions and AGB lifetimes.
In particular, according to M2K the production of $^{14}$N,
mainly of primary synthesis, is favoured at lower $Z$, and is
sistematically lower than  RV81 and HG97 results.
This latter difference can be explained considering 
the drastic reduction of TP-AGB lifetimes for the most massive
AGB models with HHB (see Fig.~\ref{fig-time}) in M2K models with respect
to RV81. 
\end{itemize}

In general,  
our expected dependence of chemical yields on metallicity
is far for being linear, and much caution should be used when
extrapolating these quantities with respect to $Z$ in chemical evolutionary
models. 
We cannot verify whether such a non-linear relation with metallicity 
was displayed also in the RV81 models, since 
just two values of metallicities were considered there.
However, the very high number of thermal pulses suffered
by stars with HBB regardless of the metallicity, according 
to the RV81 models,
might already explain the apparent lower sensitiveness of their
yields to the metallicity.

\section{Final remarks}
\label{conclu}

We would like to outline briefly the aims 
and findings of the present work. 

In the first part  
we have presented new homogeneous sets of stellar yields
ejected from low- and intermediate-mass stars, in view of providing
updated ingredients for modelling the chemical evolution
of complex stellar systems.
Thanks to the updated input physics 
employed in the calculations, and the improved treatment of both
the third dredge-up and hot-bottom burning, the present estimation 
of the stellar yields from low- and intermediate-mass stars has led to new
results and developments. 

In particular, a pronounced trend of the yields with the metallicity 
is expected. Specifically, at given stellar mass positive yields
of $^{4}$He, $^{12}$C, $^{14}$N, and $^{16}$O are larger
at decreasing metallicity.
This feature is the result of concurring factors:
at lower $Z$
both the third dredge-up and hot-bottom burning are more efficient, and
TP-AGB lifetimes are, on average, longer because of lower mass-loss rates.

Moreover, it is interesting to notice that low-mass stars 
may produce positive yields of $^{16}$O, which is brought up to
the surface by the third dredge-up. The entity of this contribution
(as well as for that of $^{12}$C)
depends crucially on the efficiency $\lambda$, number of dredge-up episodes,
and chemical composition of the convective intershell
(for this latter, see  Boothroyd \& Sackmann 1988b, and  Herwig 2000 
for different model results).   

The possible production and ejection of 
newly synthesised oxygen from low- and intermediate-mass
stars may be an interesting prediction to be tested  
through its consequences in various possible applications, 
mostly in relation to the chemical composition of planetary nebulae
(e.g. P\'equignot et al. 2000), and galactic 
chemical evolutionary models.

The second part of the paper is meant to examine several 
aspects concerning the stellar models which the chemical yields
are derived from.  To this aim,
we have analysed how the main model prescriptions
(i.e. treatment of convective boundaries, mass loss, 
analytical relations, third dredge-up parameters, treatment of 
hot-bottom burning, etc.) may affect the predictions of stellar yields. 

Finally, the third final part is dedicated to compare our new yields    
with other available data sets of large usage, in the attempt of
explaining the existing differences as the result of particular
assumptions. 
To do this, we have considered basic observational
constraints which are closely related to the stellar yields -- namely:
i) the initial-final mass relation; ii) the white dwarf mass distribution;
iii) the carbon star luminosity function; and iv) the chemical
composition of planetary nebulae -- and tested 
the capability of different models in reproducing them 
through  a direct comparison between predictions and observations.

In particular, it has been shown 
how much the choice of calibrating 
fundamental efficiency parameters (e.g. of the third dredge-up
and mass loss) in recent works has changed the 
predictions of stellar yields compared to earlier studies.

To conclude, we wish this work has somehow contributed 
to clarify a few important general points on theoretical stellar yields,
in view of stimulating an aware and critical 
usage of them.
 

\begin{acknowledgements}
I would like to thank L\'eo Girardi, Laura Portinari, and Cesare Chiosi 
for their important professional advice and support, 
and the referee, Dr. R.B.C. Henry, for his helpful remarks on this work.
This study has been financially supported by the Italian 
Ministry of University, Scientific Research and Technology (MURST)
under contract ``Formation and evolution
of Galaxies'' n.~9802192401.
\end{acknowledgements}

%
\appendix
\section{Tables of stellar yields and planetary nebulae chemical abundances}
\label{sec_app}
Tables A1 -- A12 contain the yields from low- and intermediate-mass
stars, in the form $M_{\rm y}$ (see Eq.~(\ref{defy1})), 
as a function of the initial
mass $M_{\rm i}$ and  metallicity $Z$.
All specified masses are expressed in solar units.

Concerning low-mass stars (with $M_{\rm i} \la M_{\rm HeF}$), 
we give separately the yields ejected during the RGB 
(Tables A1 -- A3) and AGB phases (Tables A4 -- A6).
The quantity, $M_{\rm TP,0}$, corresponds to the mass at the onset of the 
TP-AGB phase, which is smaller than $M_{\rm i}$ by the amount of mass lost
during the previous RGB phase, $\Delta M_{\rm ej}({\rm RGB})$.
The mass lost during the AGB is denoted with $\Delta M_{\rm ej}({\rm AGB})$.

Total yields produced by stars in the whole mass range 
($0.8 M_{\odot} \la M_{\rm i} \la 5 M_{\odot}$) are presented 
in Tables A7 -- A12. The total amount of ejected mass is denoted with
$\Delta M_{\rm ej}$.

Total yields from stars with $M_{\rm i} \ge 3.5 M_{\odot}$ are given 
in Tables A10 -- A12 for three values of the mixing-length parameter
$\alpha$, and distinguishing between the secondary (S entry) and primary 
(P entry) components in the case of the CNO elements.

Tables A13 -- A15 present the predicted abundances ratios
(He/H, C/H, N/H, O/H) in planetary nebulae,
as a function of the initial stellar mass ($M_{\rm i}$),
metallicity $Z$, and mixing-length parameter. The PNe chemical
abundances (by number, in mole gr$^{-1}$) are  
calculated by averaging the abundances in the wind ejecta
over the last stages (i.e. a time period $\tau_{\rm PN} = 3 \times 10^{4}$ yr) 
on the AGB,  weighted by the masses of the ejecta.

For the sake of simplicity, we do not consider the question on the actual
observability of PNe (which depends on both dynamical and ionisation
properties), and the fact that evolutionary timescales 
of PNe (hence the time $\tau_{\rm PN}$)
may largely vary according to the mass of the progenitor star.
These points deserve a more complex study, which is currently in progress
and presented in a preliminary form by Marigo et al. (1999b).

All data are available in electronic format 
at the web site address: http://pleiadi.pd.astro.it. 
\clearpage

\markboth{Appendix A: TABLES OF STELLAR YIELDS}{Appendix A: TABLES OF STELLAR YIELDS}
\begin{figure*}[b]
\resizebox{\textwidth}{!}{\includegraphics{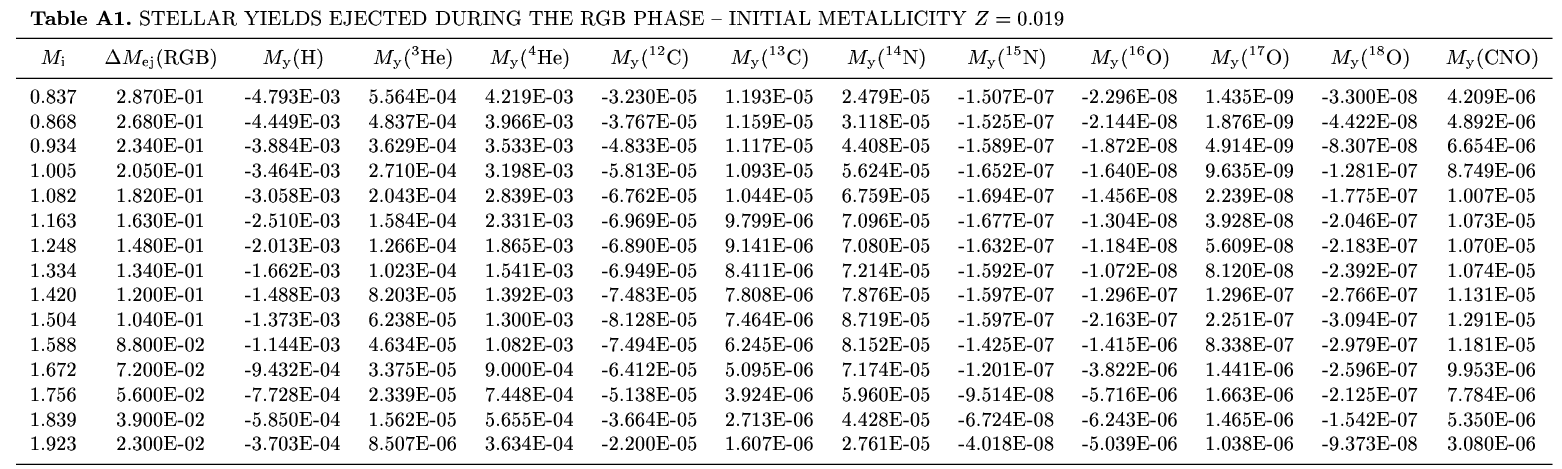}}
\end{figure*}
\begin{figure*}
\resizebox{\textwidth}{!}{\includegraphics{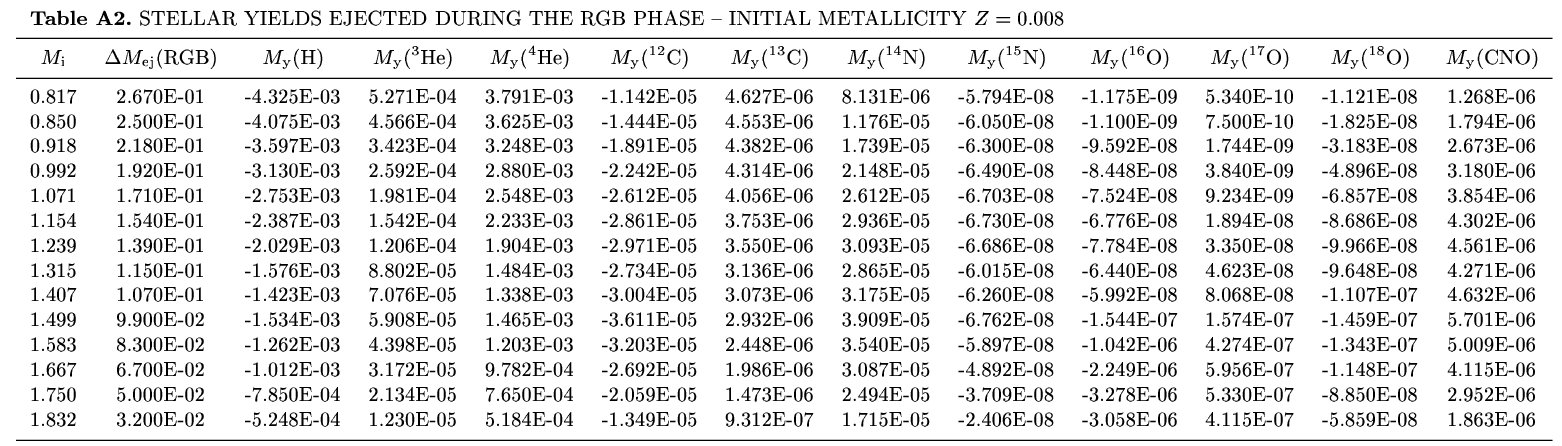}}
\end{figure*}
\begin{figure*}
\resizebox{\textwidth}{!}{\includegraphics{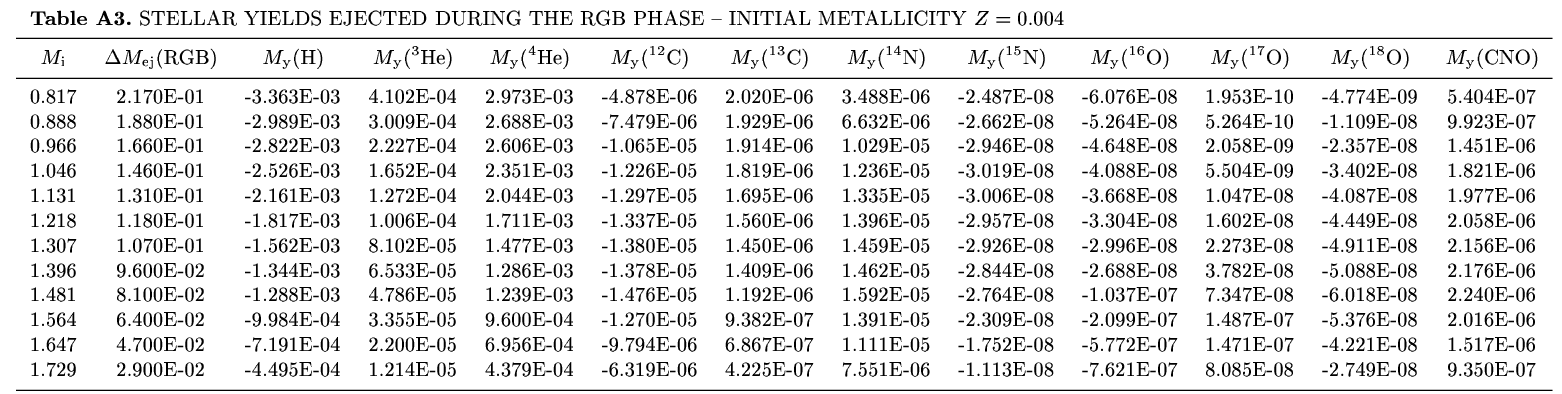}}
\end{figure*}


\begin{figure*}
\resizebox{\textwidth}{!}{\includegraphics{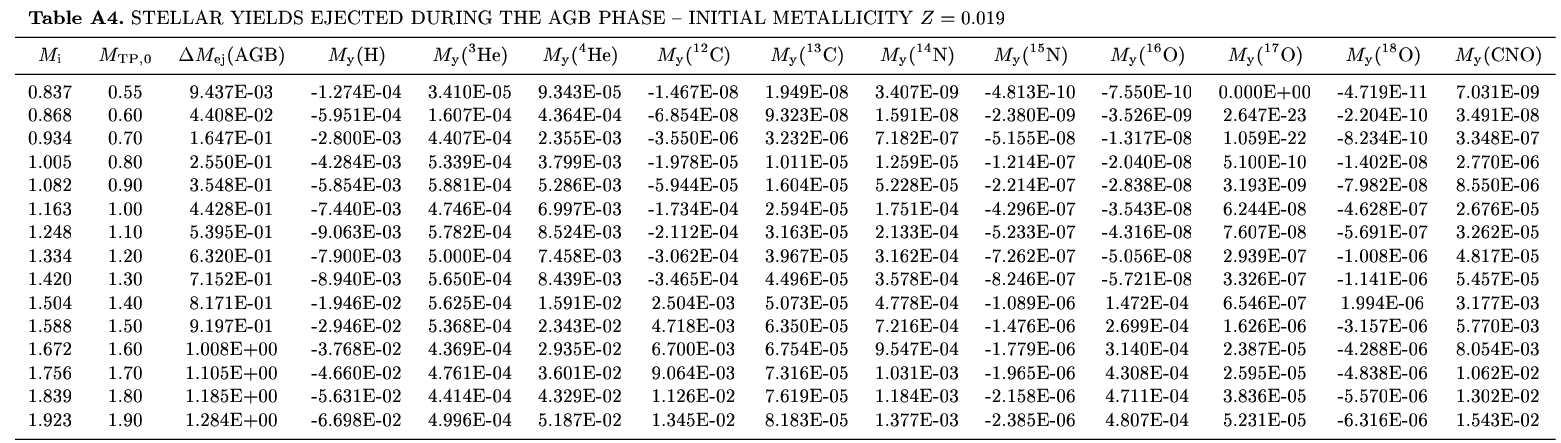}}
\end{figure*}
\begin{figure*}
\resizebox{\textwidth}{!}{\includegraphics{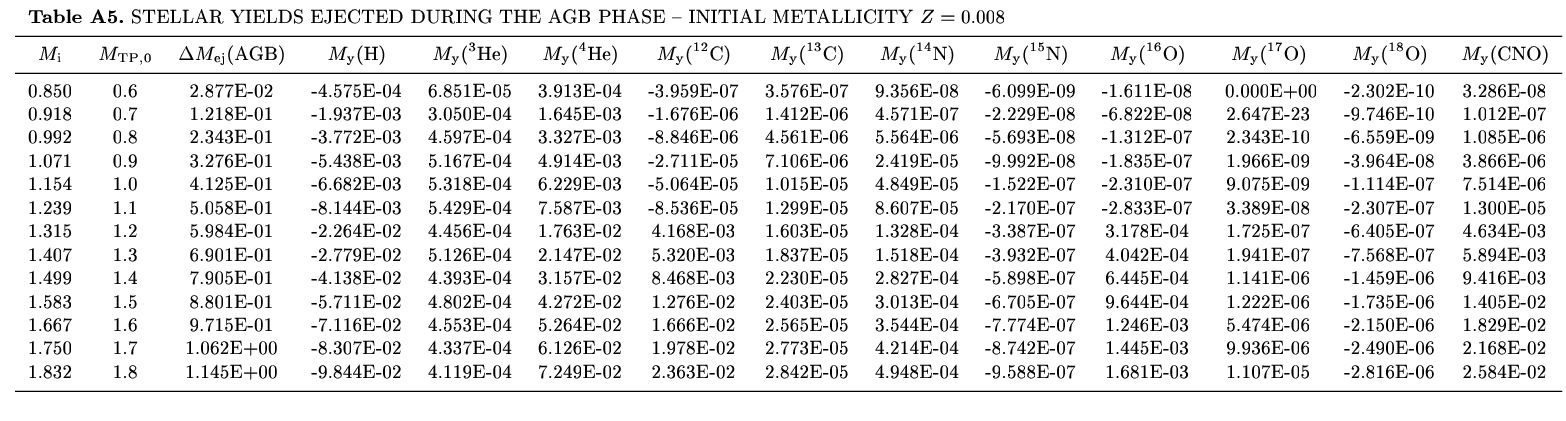}}
\end{figure*}
\begin{figure*}
\resizebox{\textwidth}{!}{\includegraphics{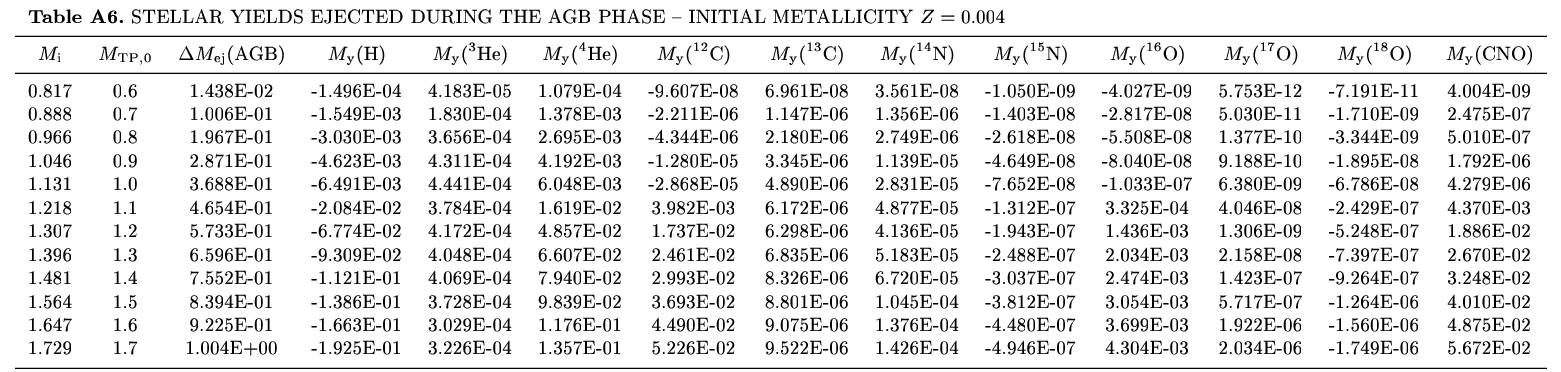}}
\end{figure*}


\begin{figure*}
\resizebox{\textwidth}{!}{\includegraphics{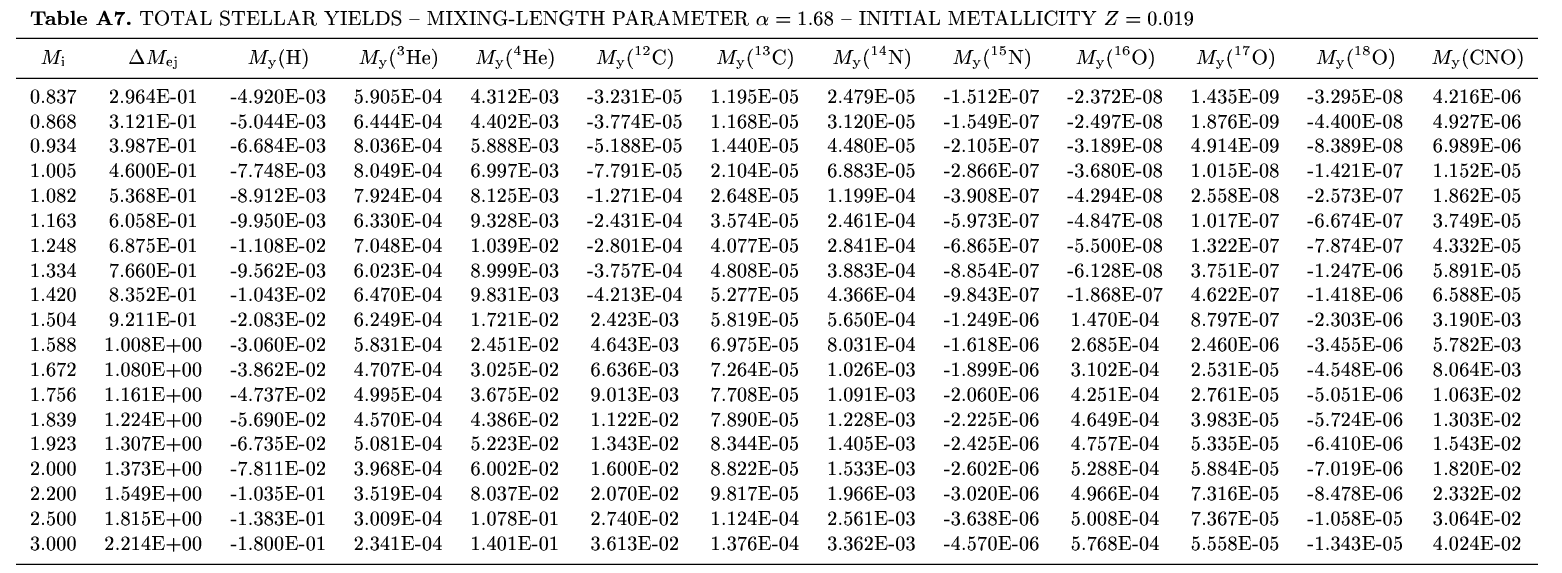}}
\end{figure*}
\begin{figure*}
\resizebox{\textwidth}{!}{\includegraphics{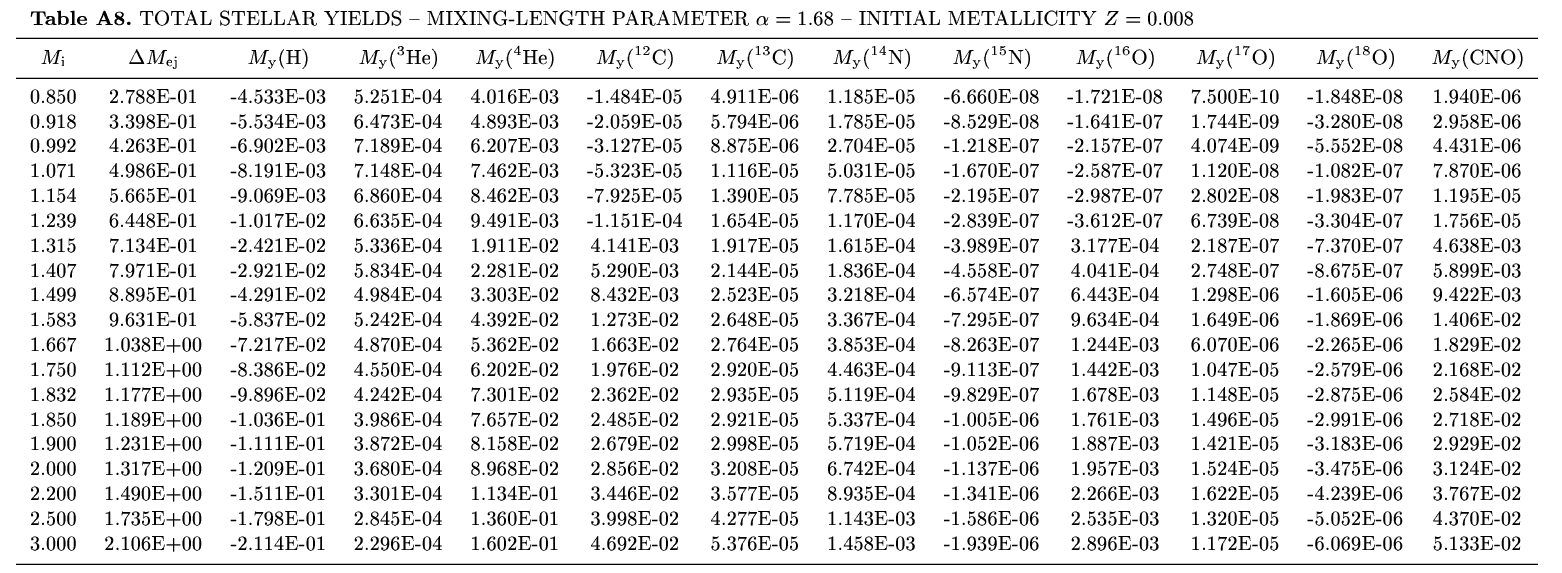}}
\end{figure*}
\begin{figure*}
\resizebox{\textwidth}{!}{\includegraphics{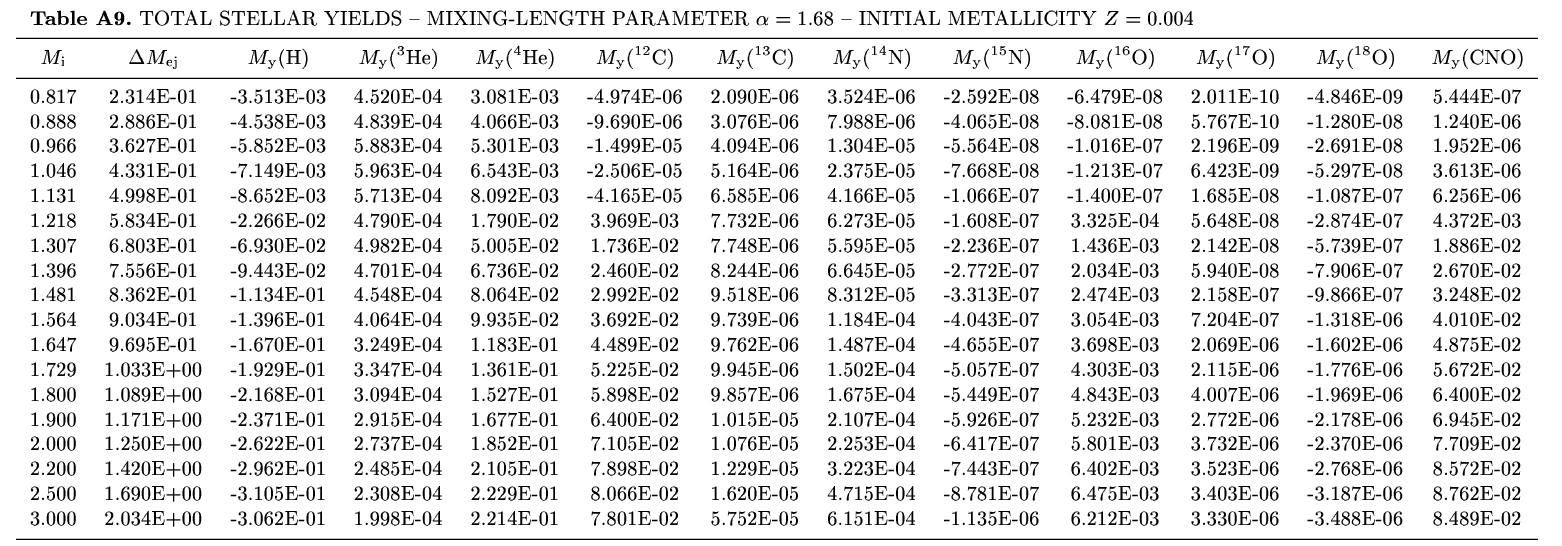}}
\end{figure*}



\begin{figure*}
\resizebox{\textwidth}{!}{\includegraphics{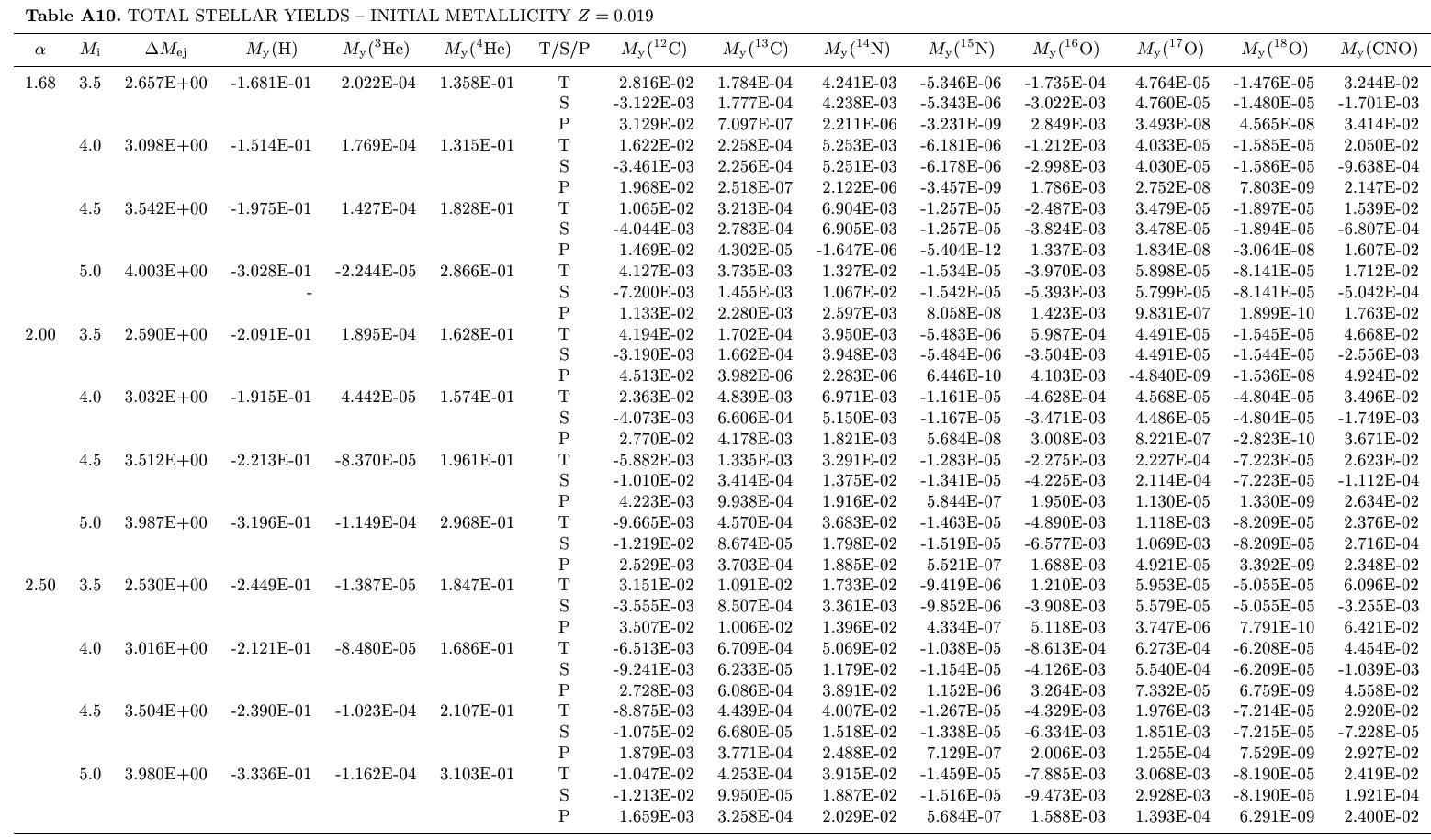}}
\end{figure*}
\begin{figure*}
\resizebox{\textwidth}{!}{\includegraphics{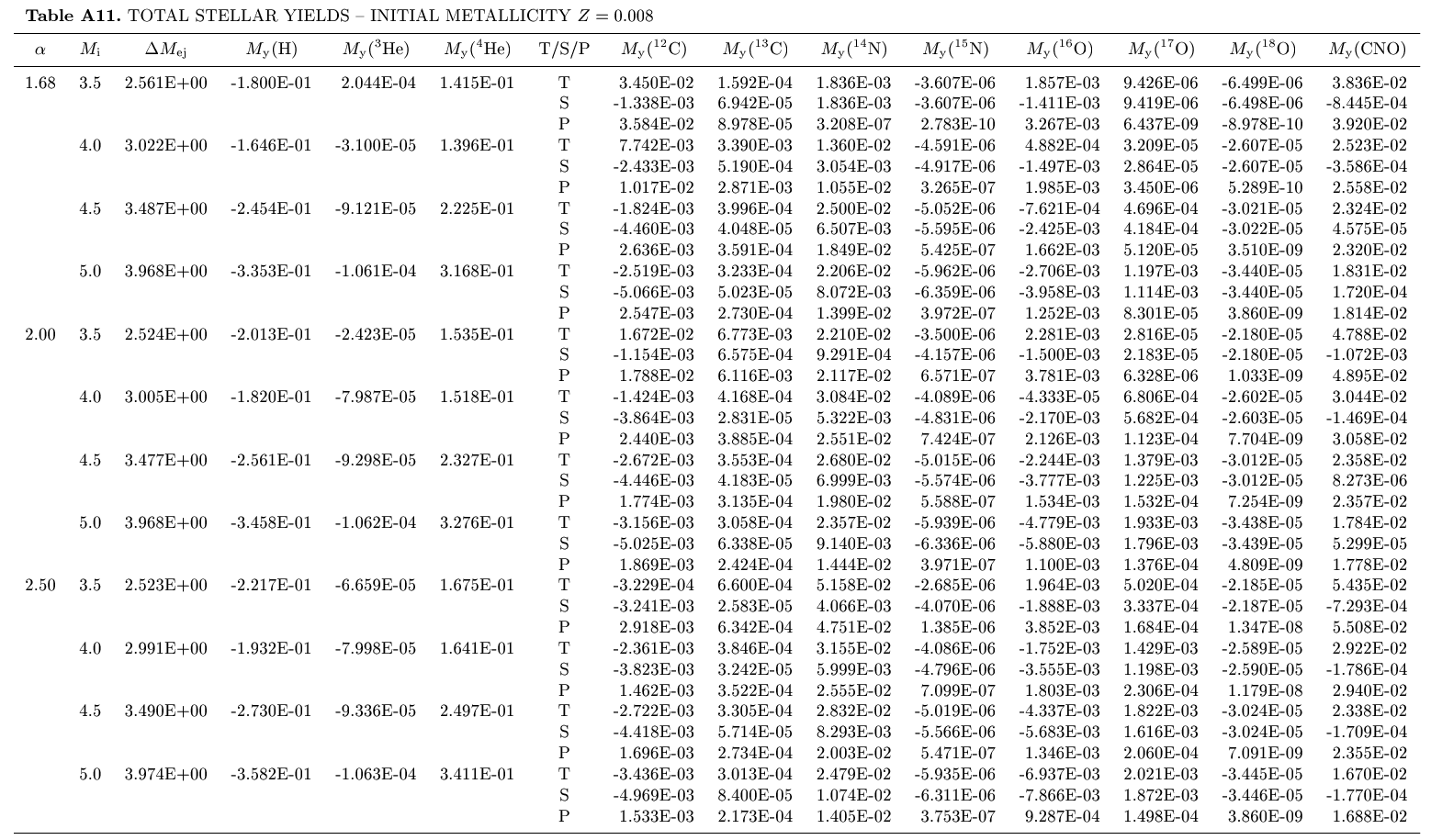}}
\end{figure*}
\begin{figure*}
\resizebox{\textwidth}{!}{\includegraphics{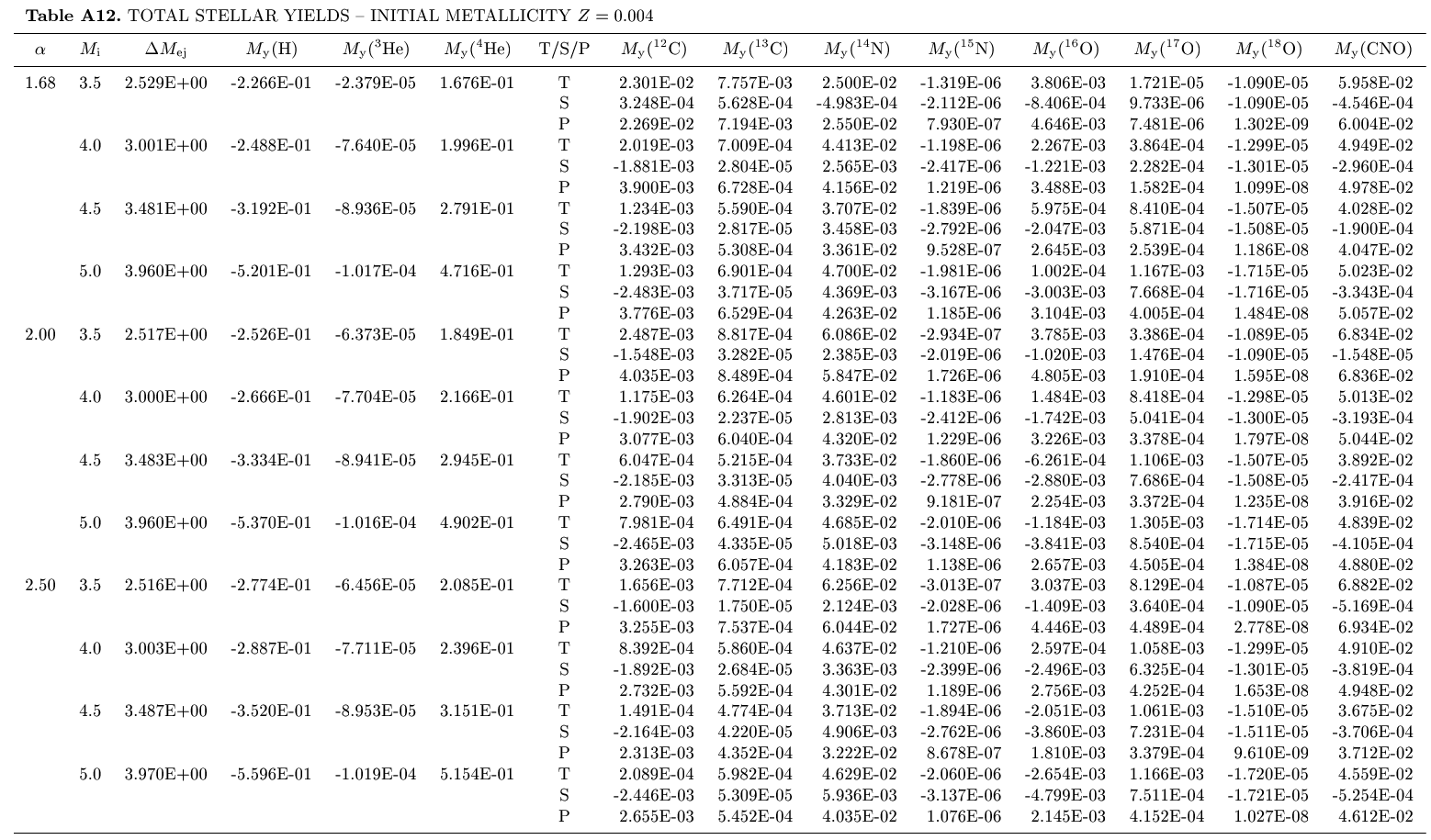}}
\end{figure*}

\clearpage
\begin{figure*}
\begin{minipage}{0.45\textwidth}
\resizebox{\hsize}{!}{\includegraphics{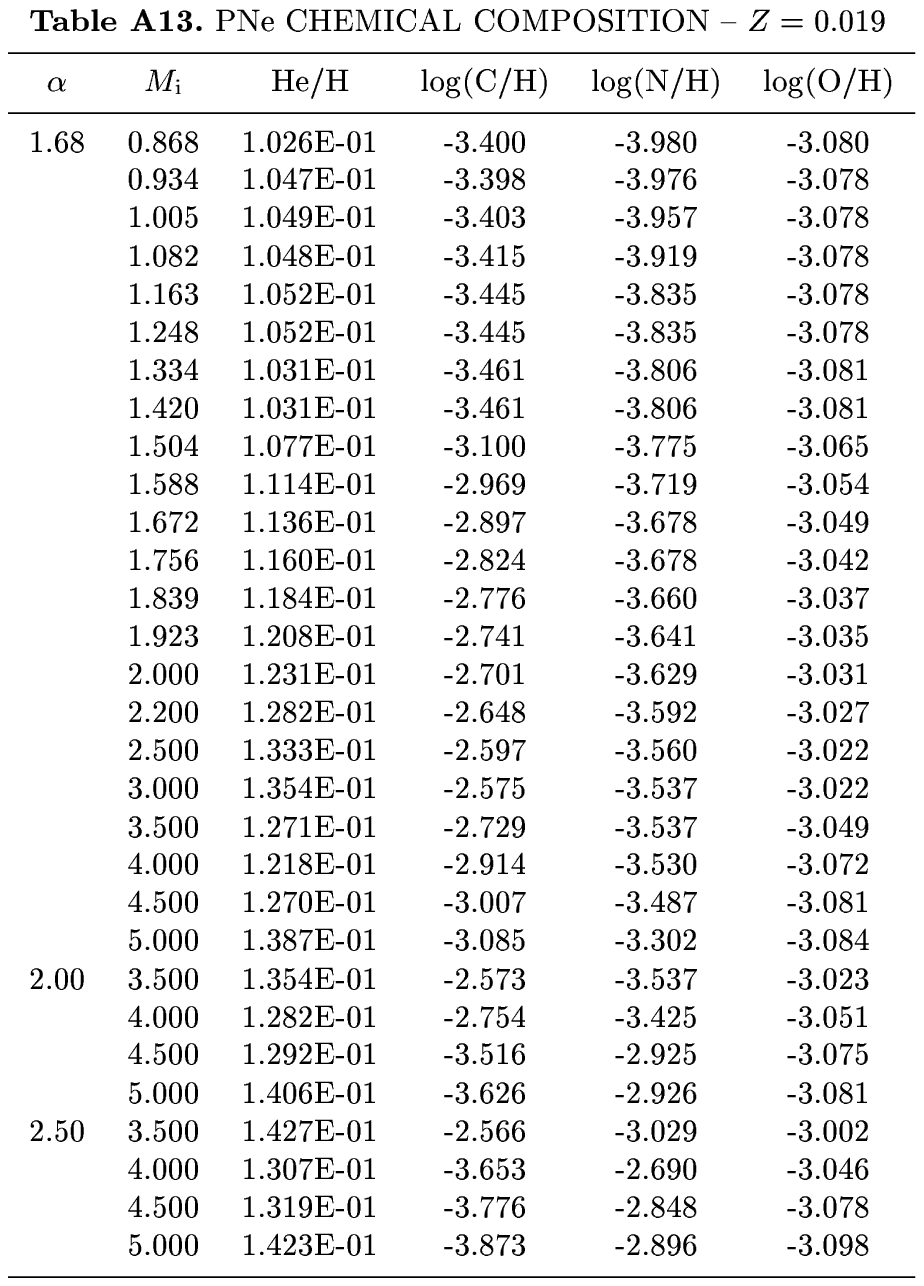}}
\end{minipage}
\hfill 
\begin{minipage}{0.45\textwidth}
\resizebox{\hsize}{!}{\includegraphics{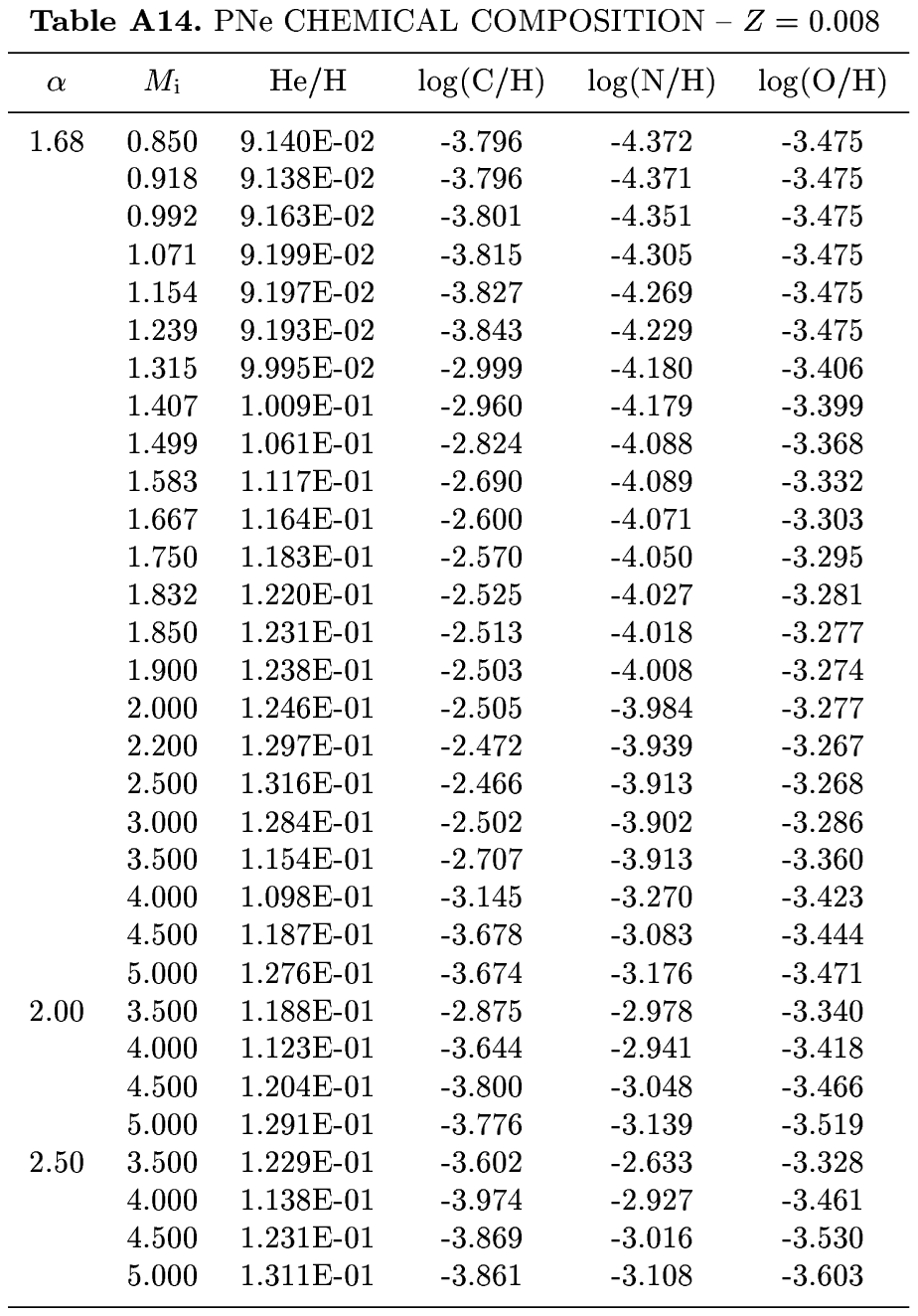}}
\end{minipage}
\begin{minipage}{0.45\textwidth}
\vspace{1truecm}
\resizebox{\hsize}{!}{\includegraphics{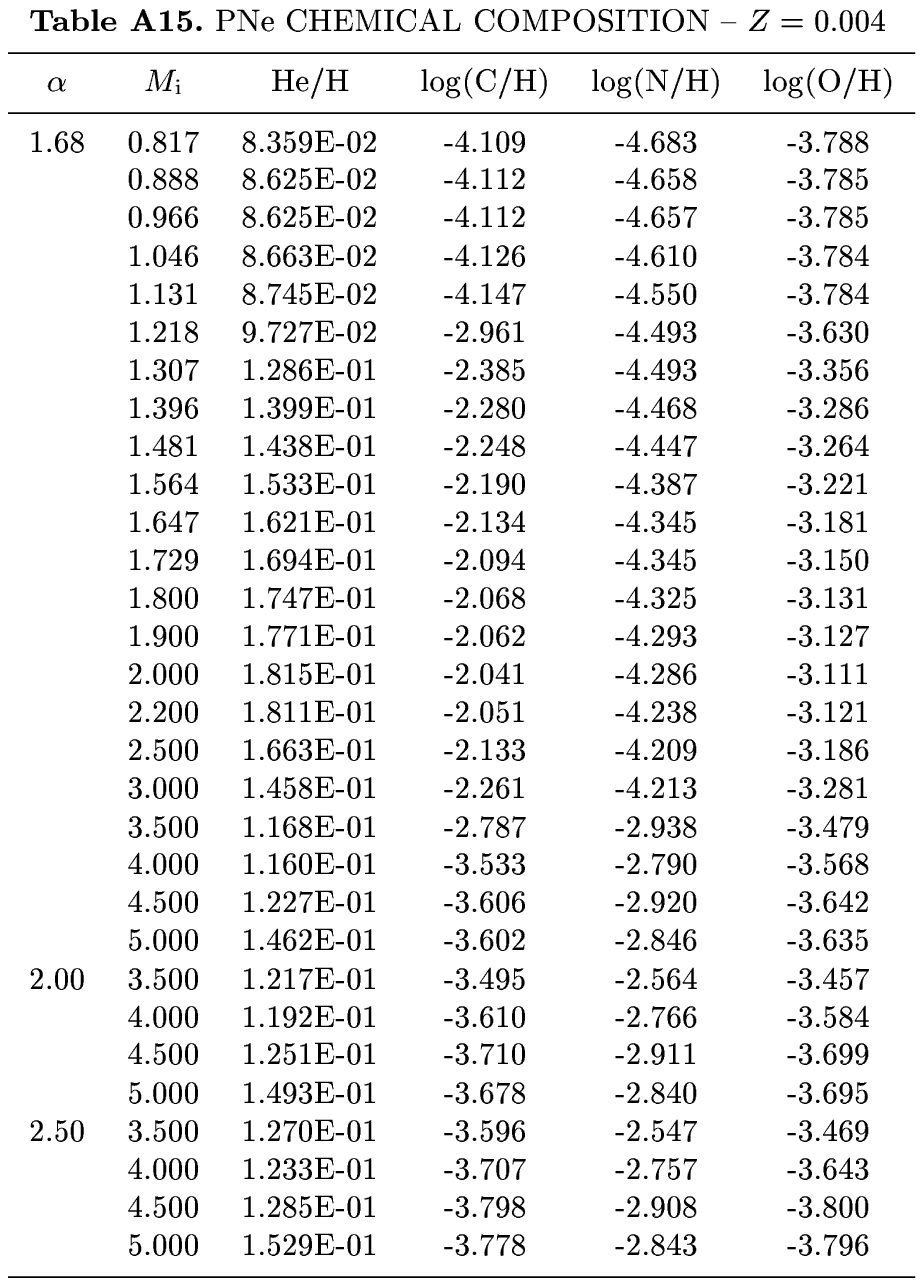}}
\end{minipage}
\hfill 
\end{figure*}


\begin{thebibliography}{}
%
\bibitem{} Alongi M., Bertelli G., Bressan A., Chiosi C., Fagotto F., 
        Greggio L., Nasi E., 1993, A\&AS 97, 851
\bibitem{} Baud B., Habing H.J., 1983, A\&A 127, 73
\bibitem{} Bergeron P., Saffer R.A., Liebert J., 1992, ApJ 394, 228
\bibitem{} Bl\"ocker T., 1995, A\&A 297, 727 (B95)
\bibitem{} Bl\"ocker T., Sch\"onberner D., 1991, A\&A 244, L43
\bibitem{} Boothroyd A.I., Sackmann I.-J., 1988a, ApJ 328, 641
\bibitem{} Boothroyd A.I., Sackmann I.-J., 1988b, ApJ 328, 653
\bibitem{} Boothroyd A.I., Sackmann I.-J., 1999, ApJ 510, 232 
\bibitem{} Bowen G.H., 1988, ApJ 329, 844  
\bibitem{} Bragaglia A., Renzini A., Bergeron P., 1995, ApJ 443, 735
\bibitem{} Caughlan G.R., Fowler W.A., 1988, Atomic Data Nucl.\ 
           Data Tables 40, 283
\bibitem{} Charbonnel C., 1995, ApJ 453, L41-L44
\bibitem{} Chiosi C., Bertelli G., Bressan A., 1992, ARA\&A 30, 305
\bibitem{} Costa E., Frogel J.A., 1996, AJ 112, 2607
\bibitem{} Crotts, A.P.S., Bechtold J., Fang Y., Duncan R.C., 1994,
      Bull. American Astron. Soc. 185, 1807
\bibitem{} Dominguez I., Chieffi A., Limongi M., et al., 1999, ApJ 524, 226
\bibitem{} Fleischer A.J., Gauger A., Sedlmayr E., 1992, A\&A 266, 321
\bibitem{} Forestini M., Charbonnel C., 1997, A\&AS 123, 241
\bibitem{} Garcia-Berro E., Ritossa C., Iben I., 1997, ApJ 485, 765
\bibitem{} Girardi L., Bressan A., Bertelli G., Chiosi C., 2000, A\&AS 141, 371
\bibitem{} Graboske H.C., de Witt H.E., Grossman A.S., Cooper M.S., 1973,
        AJ 181, 457
\bibitem{} Gratton R.G., Sneden C., Carretta E., Bragaglia A., 2000,
	   A\&A 354, 169 
\bibitem{} Groenewegen M.A.T., de Jong T., 1993, A\&A 267, 410
\bibitem{} Groenewegen M.A.T., de Jong T., 1994, A\&A 282, 115
\bibitem{} Henry R.B.C., Kwitter K.B., Bates J.A., 2000a, ApJ, 531, 928
\bibitem{} Henry R.B.C., Edmunds M.G., K\"oppen J., 2000b, ApJ, 541, 660
\bibitem{} Herwig F., 1996, in Stellar Evolution: 
        What Should Be Done, 32nd Li\`ege Int.\ Astrophys.\ Coll.,
        eds.\ A.\ Noels et al., p.\ 441
\bibitem{}  Herwig F., Bl\"ocker T., Sch\"onberner D., El Eid M., 1997, 
	A\&A 324, L81
\bibitem{} Herwig F., 2000, A\&A 360, 952
\bibitem{} van den Hoek L.B., Groenewegen M.A.T., 1997, A\&AS 123, 305 (HG97)
\bibitem{} Iben I., 1981, ApJ 246, 278
\bibitem{} Iben I., Renzini A., 1983, ARA\&A 21, 27
\bibitem{} Iben I., Truran J.W., 1978, ApJ 220, 980 (IT78)
\bibitem{} Jeffries R.D., 1997, MNRAS 288, 585
\bibitem{} Jeffries R.D., James D.J., Thurston M.R., 1998, MNRAS 300, 550
\bibitem{} Kingsburgh R.L., Barlow M.J., 1994, MNRAS 271, 257
\bibitem{} Koester D., Reimers D., 1996, A\&A 313, 810
\bibitem{} Lu, L.. 1991, ApJ 379, 99
\bibitem{} Lu, L., Sargent W.L.W., Barlow T.A., 1998, AJ 115, 55
\bibitem{} Maeder A., 1983, A\&A 120, 113
\bibitem{} Maeder A., Meynet G., 2000, ARA\&A 38, in press
\bibitem{} Marigo P., 1998a, PhD Thesis, University of Padova 
\bibitem{} Marigo P., 1998b, A\&A 340, 463
\bibitem{} Marigo P., Bressan A., Chiosi C., 1996, A\&A 313, 545
\bibitem{} Marigo P., Bressan A., Chiosi C., 1998, A\&A 331, 564
\bibitem{} Marigo P., Girardi L., Bressan A., 1999a, A\&A 344, 123
\bibitem{} Marigo P., Weiss A., Groenewegen M.A.T., Girardi L., 
        1999b, in From extrasolar planets to Cosmology: 
        The VLT Opening Symposium, Proceedings ESO/MPA 
        Parallel Workshop 2: Star-Way to the Universe, 
        held in Antofagasta (Chile), March 1999, Eds. Bergeron J., 
        Renzini A., p.~248-251
\bibitem{} Peimbert M., 1978, in IAU Symposium 76, Planetary Nebulae, 
           ed. Y. Terzian, Dordrecht: Reidel, p.~215 
\bibitem{} P\`equignot D., Walsh J.R., Zijlstra A.A., Dudziak G., 
	   2000, A\&A 361, L1
\bibitem{} Pettini M., Lipman K., Hunstead R.W., 1995, ApJ 451, 100
\bibitem{} Portinari L., Chiosi C., Bressan A., 1998, A\&A 334, 505
\bibitem{} Prantzos, N., Vangioni-Flam E., Casse M., 1994, 
           RAS Canada J. 88, 356
\bibitem{} Rebeirot E., Azzopardi M., Westerlund B.E., 1993, A\&AS 97, 603
\bibitem{} Reid, N., 1996, AJ 111 (5), 2000
\bibitem{} Reimers D., 1975, Mem.\ Soc.\ R.\ Sci.\ Li\`ege, 
        ser.\ 6, vol.\ 8, p.\ 369 
\bibitem{} Renzini A., Voli M., 1981, A\&A 94, 175 (RV81)
\bibitem{} Salpeter E.E., 1955, ApJ 121, 161
\bibitem{} Tinsley B.M., 1980, Fundam.\ of Cosmic Phys.\ 5, 287
\bibitem{} Tytler D., Fan X.-M., 1994, ApJ 424, L87
\bibitem{} Vassiliadis E., Wood P.R, 1993, ApJ 413, 641
\bibitem{} Vila-Costas M.B., Edmunds M.G., 1993, MNRAS 265, 199
\bibitem{} Wagenhuber J., Groenewegen M.A.T., 1998, A\&A 340, 183
\bibitem{} Wasserburg G.J, Boothroyd A.I., Sackmann I.-J., 1995, ApJ 447, L37
\bibitem{} Weidemann V., 1987, A\&A 188, 74 
\bibitem{} Weidemann V., 1997, in Advances in Stellar Evolution, 
           Proceedings of the Workshop Stellar Ecology, 
           Cambridge University Press, p.\ 169
\bibitem{} Weiss A., Denissenkov P.A., Charbonnel A., 2000, A\&A 355, 299 
\bibitem{} Woosley S.E., Weaver T.A., 1982, in Supernovae: 
	   a Survey of Current Research, Reidel, Dordrecht, 
            eds. Rees M. J., Stoneham R. J., p.79
\bibitem{} Wood, P.R., 1981, in: Physical processes in red giants, Proc. of
        the Second Workshop, Erice, Italy (September 3-13, 1980), Dordrecht,
        D. Reidel Publishing Co., p. 135-139
%
\end{thebibliography}
\end{document}